\documentclass[journal=jctce,manuscript=article,layout=traditional]{achemso}

\usepackage{graphicx}
\usepackage{dcolumn}
\usepackage{bm}
\usepackage{rotating}
\usepackage{multirow}
\usepackage[table]{xcolor}
\usepackage[english]{babel}
\usepackage{amssymb} %maths
\usepackage{amsmath} %maths
\usepackage[utf8]{inputenc} %useful to type directly diacritic characters
\usepackage{siunitx}
\usepackage{tabularx}% added for table design
\usepackage{subfigure}
\usepackage{lipsum}
\usepackage{hyperref}
\usepackage{wrapfig}
\usepackage{ulem}

\SectionNumbersOn

\newcommand{\be}{\begin{equation}}
\newcommand{\ee}{\end{equation}}
\newcommand{\bea}{\begin{eqnarray}}
\newcommand{\eea}{\end{eqnarray}}
\newcommand{\nn} {\nonumber}

\newcommand{\bG}{{\bf G}}
\newcommand{\br}{{\bf r}}

\newcommand{\bR}{{\bf R}}
\newcommand{\bT}{{\bf T}}
\newcommand{\bq}{{\bf q}}

\newcommand{\bk}{{\bf k}}
\newcommand{\bzero}{{\mathbf{0}}}

\newcommand{\bp}{{\bf p}}

\newcommand{\me}[3]{\langle #1 | #2 | #3\rangle}

\def\QE{\textsc{Quantum ESPRESSO}\,}

\title{Koopmans spectral functionals in periodic-boundary conditions}
\author{Nicola Colonna}
\email{nicola.colonna@psi.ch}\affiliation{Laboratory for Neutron Scattering and Imaging, Paul Scherrer Institute, 5232 Villigen PSI, Switzerland}
\alsoaffiliation{National Center for Computational Design and Discovery of Novel Materials (MARVEL), \'{E}cole Polytechnique F\'{e}d\'{e}rale de Lausanne, 1015 Lausanne, Switzerland}
\author{Riccardo De Gennaro}
\affiliation{National Center for Computational Design and Discovery of Novel Materials (MARVEL), \'{E}cole Polytechnique F\'{e}d\'{e}rale de Lausanne, 1015 Lausanne, Switzerland}
\alsoaffiliation{Theory and Simulations of Materials (THEOS), \'{E}cole Polytechnique F\'{e}d\'{e}rale de Lausanne, 1015 Lausanne, Switzerland}
\author{Edward Linscott}
\affiliation{National Center for Computational Design and Discovery of Novel Materials (MARVEL), \'{E}cole Polytechnique F\'{e}d\'{e}rale de Lausanne, 1015 Lausanne, Switzerland}
\alsoaffiliation{Theory and Simulations of Materials (THEOS), \'{E}cole Polytechnique F\'{e}d\'{e}rale de Lausanne, 1015 Lausanne, Switzerland}
\author{Nicola Marzari}
\affiliation{National Center for Computational Design and Discovery of Novel Materials (MARVEL), \'{E}cole Polytechnique F\'{e}d\'{e}rale de Lausanne, 1015 Lausanne, Switzerland}
\alsoaffiliation{Theory and Simulations of Materials (THEOS), \'{E}cole Polytechnique F\'{e}d\'{e}rale de Lausanne, 1015 Lausanne, Switzerland}
\alsoaffiliation{Laboratory for Materials Simulations, Paul Scherrer Institut, 5232 Villigen PSI, Switzerland}

\date{\today}

\begin{document}
%%%%% TOC 
%\begin{figure}
%\caption*{TOC}\label{fig:TOC}
%\includegraphics[width=12.5cm]{Figs/TOC_horizontal.pdf}
%\end{figure} 

\begin{abstract}
Koopmans spectral functionals aim to describe simultaneously ground state properties and charged excitations of
atoms, molecules, nanostructures and periodic crystals. This is achieved by augmenting standard density 
functionals with simple but physically motivated orbital-density-dependent corrections. These corrections 
act on a set of localized orbitals that, in periodic systems, resemble maximally localized Wannier 
functions. At variance with the original, direct supercell implementation [{\it Phys. Rev. X} {\bf 8}, 
021051 (2018)], we discuss here i) the complex but efficient formalism required for a periodic-boundary 
code using explicit Brillouin zone sampling, and ii) the calculation of the screened Koopmans corrections 
with density-functional perturbation theory. In addition to delivering improved scaling with 
system size, the present development makes the calculation of band structures with Koopmans functionals 
straightforward. The implementation in the open-source \QE distribution and the application to 
prototypical insulating and semiconducting systems are presented and discussed.
\end{abstract}

\maketitle

\section{Introduction}
Electronic-structure simulations have a profound impact on many scientific fields, from 
condensed-matter physics to chemistry, materials science, and 
engineering~\cite{marzari_electronic-structure_2021}. One of the main reasons for this is the accuracy and 
efficiency of Kohn-Sham density-functional theory~\cite{hohenberg_inhomogeneous_1964, kohn_self-consistent_1965} 
(KS-DFT), together with the availability of robust computational tools that implement and make available 
these fundamental theoretical developments. Nevertheless, exact KS-DFT can only describe (if the exact 
functional were known) the total energy of a system, including its static derivatives (or the expectation 
value of any local single-particle operator), precluding any access to spectroscopic information, except 
for the position of the highest occupied orbital~\cite{levy_exact_1984,almbladh_exact_1985,perdew_comment_1997} 
(HOMO) (see also Ref.~\citenum{chong_interpretation_2002} and Ref.~\citenum{pederson_densityfunctional_1985} and
references therein for an in-depth discussion about the connection between KS eigenvalues and vertical 
ionization potentials). 

While the access to charge-neutral excitations can be achieved by extending the formalism to the time 
domain~\cite{runge_density-functional_1984}, charged excitations --- revealed in direct and inverse 
photoemission experiments --- are outside the realm of the theory. Accurate first-principles predictions
of band gaps, photoemission spectra, and band structures require more advanced approaches, most often 
based on Green’s function theory~\cite{onida_electronic_2002}. For example, in solids, the so-called 
GW approximation~\cite{hedin_new_1965} is considered a good compromise between accuracy and computational cost. 
Nevertheless, these high-level methods are often limited in system size and complexity, due to their 
computational cost and numerical complexity. Despite many efforts dedicated to improving efficiency of 
Green's function methods~\cite{umari_optimal_2009, umari_gw_2010, giustino_gw_2010, neuhauser_breaking_2014, govoni_large_2015, wilhelm_toward_2018, vlcek_swift_2018, umari_fully_2022},
simpler methods based on Kohn-Sham density-functional theory (KS-DFT), possibly including some fraction
of non-local exchange~\cite{becke_new_1993}, are still frequently employed to approximately evaluate the
spectral properties of nanostructures, interfaces, and solids. 
In this respect, Koopmans-compliant (KC) 
functionals~\cite{dabo_non-koopmans_2009, dabo_koopmans_2010, dabo_piecewise_2014, borghi_koopmans-compliant_2014, borghi_variational_2015, nguyen_koopmans-compliant_2018} 
have been introduced to bridge the gap between KS-DFT and Green's function theory~\cite{ferretti_bridging_2014, colonna_koopmans-compliant_2019}. 
KC functionals retain the advantages of a functional formulation by enforcing physically motivated 
conditions to approximate density functionals. 
In particular, the exact condition of the piecewise linearity (PWL) of the total energy as a function
of the total number of electrons~\cite{perdew_density-functional_1982}, or equivalently of the occupation
of the HOMO, is extended to the entire manifold, leading to a \textit{generalized PWL} of the energy 
as a function of each orbital occupation~\cite{dabo_koopmans_2010, borghi_koopmans-compliant_2014}.
In KS-DFT the deviation from PWL has been suggested~\cite{cococcioni_linear_2005,kulik_density_2006,mori-sanchez_many-electron_2006, cohen_insights_2008,mori-sanchez_localization_2008} 
as a definition of electronic self-interaction errors (SIEs)~\cite{perdew_self-interaction_1981}, and 
in recently developed functionals, such as DFT-corrected~\cite{zheng_improving_2011,zheng_nonempirical_2013, kraisler_piecewise_2013,kraisler_fundamental_2014, li_local_2015,gorling_exchange-correlation_2015,li_localized_2018, mei_exact_2021}, range-separated~\cite{stein_fundamental_2010,kronik_excitation_2012,refaely-abramson_quasiparticle_2012} 
or dielectric-dependent hybrid functionals~\cite{shimazaki_band_2008, marques_density-based_2011,skone_self-consistent_2014,brawand_generalization_2016},
it has been recognized as a critical feature to address. The generalized PWL of KC functionals leads
to beyond-DFT orbital-density dependent functionals, with enough flexibility to correctly describe 
both ground states and charged excitations~\cite{dabo_donor_2012, nguyen_first-principles_2015, nguyen_first-principles_2016, elliott_koopmans_2019, colonna_koopmans-compliant_2019, de_almeida_electronic_2021}.
In fact, while ground-state energies are typically very close or exactly identical to those of the
``base'' functional~\cite{borghi_koopmans-compliant_2014}, some of us argued that, for spectral 
properties, the orbital-dependent KC potentials act as a quasiparticle approximation to the spectral
potential~\cite{ferretti_bridging_2014} (that is, the local and frequency-dependent potential sufficient
to correctly describe the local spectral density $\rho(\br, \omega)$~\cite{gatti_transforming_2007, vanzini_dynamical_2017, vanzini_spectroscopy_2018}).

Beside the core concept of generalized PWL, KC functionals are characterized by two other features: 
i) the correct description of the screening and relaxation effects that take place when an electron 
is added/removed from the system~\cite{zhang_orbital_2015, colonna_screening_2018}, and ii) the 
localization of the ``variational'' orbitals, i.e. those that minimize the KC energy. This last
feature is key to obtaining meaningful and accurate results in extended or periodic 
systems~\cite{nguyen_koopmans-compliant_2018,de_almeida_electronic_2021}, but at the same time poses
some challenges since the localized nature of the variational orbitals apparently breaks the translational symmetry. 
Nevertheless, thanks to the Wannier-like character of the variational orbitals, the Bloch symmetry
is still preserved and it is possible to describe the electronic energies with a band structure 
picture~\cite{de_gennaro_blochs_2021}. While a general method to unfold and interpolate the electronic
bands from $\Gamma$-point-only calculations can be employed~\cite{de_gennaro_blochs_2021}, in this
work we describe how to exploit the Wannier-like character of the variational orbitals to recast 
the Koopmans corrections as integrals over the Brillouin zone of the primitive cell. This leads
to a formalism suitable for a periodic-boundary implementation and to the natural and 
straightforward recovery of band structures. Moreover, we show how the evaluation of 
the screened KC corrections can be recast into a linear-response problem suitable for an efficient
implementation based on density-functional perturbation theory. The advantage with respect to a 
$\Gamma$-point calculation with unfolding is a much reduced computational cost and complexity.
In the rest of the paper we will describe the details of such a formalism, which leads to a 
transparent and efficient implementation of Koopmans functionals for periodic systems. 

\section{Koopmans spectral functionals}
\label{sec:KC_general}
We review in this section the theory of KC functionals. In Sec.~\ref{sec:basic_features} 
we describe the basic features of the KC functionals; in Sec.~\ref{sec:theory_technical} 
we detail, for the interested readers, more practical and technical aspects of the method. Finally in 
Sec.~\ref{sec:minimization} we describe the general strategy to minimize the KC functionals and the 
assumptions made in this work to simplify the formalism. For a complete and exhaustive description 
of the theory we also refer the reader to previous publications~\cite{borghi_koopmans-compliant_2014, nguyen_koopmans-compliant_2018}.

\subsection{Core concepts of the theory}
\label{sec:basic_features}
\noindent
{\bf Linearization:} The basic idea KC functionals stand on is that of enforcing a generalized PWL condition of the total 
energy as a function of the fractional occupation of any orbital in the system:
\begin{equation}
    \frac{dE^{\rm KC}}{d f_i} = \langle \phi_i | \hat{h}^{\rm KC} | \phi_i \rangle =  \mathrm{constant},
    \label{eq:kc-cond}
\end{equation}
where $f_i$ is the occupation number of the $i$-th orbital $\phi_i$ and $\hat{h}^{\rm KC}$ the Koopmans-compliant Hamiltonian. Under such condition the total energy remains unchanged when an electron is, e.g., extracted ($f_i$
goes from $1$ to $0$) from the system, in analogy to what happen in a photoemission experiment.
The generalized PWL condition in Eq.~\ref{eq:kc-cond} can be achieved by simply augmenting any approximate density functional $E^{\rm DFT}$ with a set of orbital-density-dependent corrections $\{\Pi_{i}\}$ (one for each orbital $\phi_i$):
\begin{equation}\label{eq:KC_lin}
 E^{\rm KC}[\rho, \{\rho_i\}] = E^{\rm DFT}[\rho] + \sum_i {\Pi_i}[\rho, \rho_i]
\end{equation}
where $\rho(\br) = \sum_i \rho_i(\br)$ and $ \rho_i(\br) = f_i n_i(\br) = f_i |\phi_i(\br)|^2$  
are the total and orbital densities, respectively. The corrective 
term removes from the approximate DFT energy the contribution that is non-linear in the orbital
occupation $f_i$ and adds in its place a linear Koopmans’s term in such a way to satisfy the KC condition in Eq.~\ref{eq:kc-cond}. Depending on the slope of this
linear term, different KC flavors can be defined~\cite{borghi_koopmans-compliant_2014}; in this
work we focus on the Koopmans-Integral (KI) approach, where the 
linear term is given by the integral between occupations 0 and 1 of the 
expectation value of the DFT Hamiltonian on the orbital at hand:
\begin{align}\label{eq:KI_integral}
  {\Pi_i^{\rm KI}}[\rho, \rho_i] = 
  - \int_0^{f_i} ds \langle \phi_i | H^{\rm DFT}(s) | \phi_i \rangle \nonumber \\
   + f_i \int_0^{1} ds \langle \phi_i | H^{\rm DFT}(s) | \phi_i \rangle.  
\end{align}
In the expression above, the first line removes the non-linear behaviour of the underlying DFT energy
functional and the second one replaces it with a linear term, i.e. proportional to 
$f_i$. Neglecting orbital relaxations, i.e. the dependence of the orbital 
$\phi_i$ on the occupation numbers, and recalling that $\langle \phi_i | H^{\rm DFT}(s) | \phi_i \rangle = dE^{\rm DFT}/df_i$, 
the explicit expression for the ``bare'' or ``unrelaxed'' KI correction becomes: 
\begin{align}\label{eq:KI} 
     \Pi^{\rm KI}_i &=  E^{\rm DFT}_{\rm Hxc} [\rho-\rho_i] -E^{\rm DFT}_{\rm Hxc}[\rho] \nn \\ 
     & +f_i \Big[ E^{\rm DFT}_{\rm Hxc}[\rho-\rho_i+n_i] -E^{\rm DFT}_{\rm Hxc}[\rho-\rho_i] \Big],
\end{align}
where $E^{\rm DFT}_{\rm Hxc}$ is the Hartree, exchange and correlation energy at the ``base'' DFT level.
Interestingly, the KI functional is identical to the underlying DFT functional at integer 
occupation numbers ($f_i =0$ or $f_i =1$) and thus it preserves exactly the potential energy 
surface of the base functional. However, its value at fractional occupations differs from the
base functional, and thus so do the derivatives everywhere, including at integer occupations,
and consequently the spectral properties will be different.
%~\footnote{The derivative of the
%energy with respect to the occupation $f_i$ is related to the expectation value of the 
%Hamiltonian on the orbtial $\phi_i$; e.g. $dE^{\rm DFT}/df_i= \langle \phi_i| h^{\rm DFT} | \phi_i\rangle$}. 

\noindent
{\bf Screening:} By construction, the ``unrelaxed'' KI functional is linear as a function of the occupation
number $f_i$, when orbital relaxations are neglected. This is analogous to Koopmans' theorem
in HF theory, and it is not enough to guarantee the linearity of the functional in the general
case, where each orbital will relax in response to a change in the occupation of all the others. 
To enforce the generalized Koopmans' theorem --- that is, to achieve the desired linearity in 
the presence of orbital relaxation --- a set of scalar, orbital dependent screening coefficients 
are introduced that transform the ``unrelaxed'' KI correction into a fully relaxed one:
\begin{equation}\label{eq:rKI}
    E^{\rm KI}[\rho, \{\rho_i\}] = E^{\rm DFT}[\rho] + \sum_i \alpha_i {\Pi^{\rm KI}_i}[\rho, \rho_i].
\end{equation}
The scalar coefficients $\alpha_i$ act as a compact measure of electronic screening in an orbital basis
and they are given by 
a well-defined average of the microscopic dielectric function~\cite{dabo_piecewise_2014, colonna_screening_2018}:
\begin{equation}\label{eq:alpha_lr}
     \alpha_i = \frac{d^2E^{\rm DFT}/df_i^2}{\partial ^2E^{\rm DFT}/\partial f_i^2} = \frac{ \langle n_{i} | \left[ \epsilon^{-1}f_{\rm Hxc} \right] | n_{i} \rangle}{\langle n_{i} | \left[ f_{\rm Hxc} \right] | n_{i} \rangle}
\end{equation}
where $f_{\rm Hxc}(\br, \br')$ is the Hartree and exchange-correlation kernel, i.e. the second 
derivative of the underlying DFT energy functional with respect to the total density, and $\epsilon^{-1} (\br, \br')$ is 
the static microscopic dielectric function. The different notation in the derivative at the numerator 
and denominator indicates whether orbitals relaxation is accounted for ($df_i$) or not ($\partial f_i$). 

\noindent
{\bf Localization:} Similarly to other orbital-density-dependent functionals, KC functionals can
break the invariance of the total energy with respect to unitary rotations of the occupied manifold.
This implies that the energy functional is minimized by a unique set of ``variational'' 
orbitals; contrast this with density functional theory, or any unitary invariant theory, 
where any unitary transformation of the occupied manifold would leave the energy unchanged.
These variational orbitals are typically very localized in space and closely resemble
Foster-Boys orbitals~\cite{boys_construction_1960, foster_canonical_1960} in the case of atoms and molecules or
equivalently maximally localized Wannier functions~\cite{marzari_maximally_1997, marzari_maximally_2012} (MLWF)
in the case of periodic systems. This localization is driven by a Perdew-Zunger (PZ) self-interaction-correction (SIC) term
appearing in any KC functional (see Sec.~\ref{sec:ki_kipz} for the specific case of KI),
and in particular by the self-Hartree contribution to it. This also explain the similarity 
between variational orbitals and MLWF as the maximization of the self-Hartree energy and the maximal
localization produce very similar localized representations of the electronic manifold~\cite{marzari_maximally_2012}.

We recently showed that the localization of the variational orbitals is a key feature
to get meaningful KC corrections in the thermodynamic limit of extended systems (see Fig. 1 in 
Ref.~[\citenum{nguyen_koopmans-compliant_2018}] and the related discussion). 
We note in passing that several recent strategies to address the DFT band-gap underestimation
in periodic crystals have relied on the use of some kinds of localized orbitals, ranging from
defect states~\cite{miceli_nonempirical_2018, bischoff_adjustable_2019, bischoff_nonempirical_2019}
to different types of Wannier functions~\cite{heaton_self-interaction_1983, anisimov_transition_2005, ma_using_2016, weng_wannier_2018, li_localized_2018, weng_wannierkoopmans_2020, wing_band_2021, shinde_improved_2020}. 
It is a strength of KC theory that a set of localized orbitals arises naturally from the 
energy functional minimization. Indeed, this provides a more rigorous justification of the aforementioned
approaches where the use of Wannier functions is a mere (albeit reasonable) ansatz.

\subsection{Technical aspects of the theory}
\label{sec:theory_technical}
Having described the three main pillars that KC functionals stand on, in this subsection we detail several 
technical aspects of the theory. While these details are certainly important, this subsection can be
skipped without compromising the article's message. 

\subsubsection{Canonical and variational orbitals}
As we mentioned above, KC functionals depend on the orbital densities (rather than on 
the total density as in DFT or on the density matrix as in Hartree-Fock). This can break the invariance of
the total energy with respect to unitary rotations of the occupied manifold. A striking consequence
of this feature is the fact that the variational orbitals $\{\phi_i\}$, i.e. those that minimize the energy 
functional, are different from the eigenstates or ``canonical'' orbitals, i.e. those that diagonalize the 
orbital-density-dependent Hamiltonian. This duality between canonical and variational orbital
is not a unique feature of KC functionals but it also arises in any other orbital-density-dependent
functional theories such as the well-known self-interaction-correction (SIC) scheme by Perdew and 
Zunger (PZ) and has been extensively discussed in the literature~\cite{heaton_self-interaction_1983, pederson_localdensity_1984, pederson_densityfunctional_1985,lehtola_variational_2014,vydrov_tests_2007,stengel_self-interaction_2008,hofmann_using_2012}. 
At the minimum of the energy functional, the KC Hamiltonian is defined in the basis of the variational orbitals as
\begin{equation}
    \langle \phi_j | \hat{h}_i^{\rm KC} | \phi_i \rangle = \langle \phi_j | \hat{h}^{\rm DFT}[\rho] + \hat{\mathcal{V}}_i^{\rm KC}[\rho, \rho_i] | \phi_i \rangle
    \label{eq:KC_ham}
\end{equation}
where $\mathcal{V}_i^{\rm KC}[\rho, \rho_i](\br) = \delta \sum_j \alpha_j \Pi^{\rm KC}_j[\rho,\rho_j]/\delta \rho_j(\br)$
is the orbital-density-dependent KC potential. This Hamiltonian is then diagonalized to obtain the KC eigenvalues and canonical orbitals.
These orbitals are the analogue of KS-DFT or Hartree-Fock eigenstates: they usually display the symmetry of 
the Hamiltonian and, in analogy to exact DFT~\cite{levy_exact_1984,almbladh_exact_1985,perdew_comment_1997}, 
the energy of the highest occupied canonical orbitals has been numerically proven to be related to the asymptotic
decay of the ground-state charge density~\cite{stengel_self-interaction_2008}. For these reasons, canonical orbitals
and energies are usually interpreted as Dyson orbitals and quasiparticle 
energies~\cite{pederson_densityfunctional_1985, nguyen_first-principles_2015, ferretti_bridging_2014} 
accessible, for example, via photoemission experiments.

\subsubsection{Resolving the unitary invariance of the KI functional}
\label{sec:ki_kipz}
While in the most general situation KC functionals break the invariance of
the total energy under unitary rotation of the occupied manifold, in the particular
case of KI at integer occupation number the functional is invariant under such transformation. 
(Integer occupation is the typical case of an insulating system, where the valence manifold 
is separated by the conduction manifold by a finite energy gap). Indeed,
it is easy to verify that for $f_i=0$ or $f_i=1$ the KI energy correction in 
Eq.~\ref{eq:KI_integral} vanishes and the KI functional coincides with the underlying
density functional approximation which is invariant under such transformations. 
Nevertheless the spectral properties will depend on the orbital representation 
the Koopmans Hamiltonians operate on, and it is thus important to remove this 
ambiguity; this is achieved by defining the KI functional as the limit of the KIPZ functional 
at zero PZ correction~\cite{borghi_koopmans-compliant_2014}.  
Formally, this amounts to adding an extra term $-\gamma f_i E^{\rm DFT}_{\rm Hxc}[n_i]$
to the KI correction defined in Eq.~\ref{eq:KI}. In the limit of $\gamma \rightarrow 0$ this 
extra term drives the variational orbitals toward a localized representation without
modifying the energetics. Practically, this infinitesimal term is not included in the functional;
instead, the KI variational orbitals are generated by minimizing the PZ energy of the system with
respect to unitary rotations of the canonical orbitals of the base DFT functional. The final result
is entirely equivalent to taking the $\gamma \rightarrow 0$ limit.

\subsubsection{Restriction to insulating systems}
\label{sec:limitation_metals}
The KC linearization procedure can be imposed upon both the valence and conduction states. Currently,
the only requirement is that the system under consideration needs to have a finite band gap so that
the occupation matrix can always be chosen to be block-diagonal and equal to the identity for the 
occupied manifold and zero for the empty manifold~\cite{dabo_piecewise_2014}. This limitation 
follows from the fact that currently KC corrections are well-defined for changes in the diagonal
elements of the occupation matrix. In the most general case
where a clear distinction between occupied and empty manifold is not possible, e.g. 
in the case of metallic systems, the occupation matrix will be necessarily non-diagonal 
in the localized-orbitals representation. This would in turn call for possibly more general
KC corrections to deal with such off-diagonal terms in the occupation matrix. While this
would certainly be a desirable improvement of the theory, as things currently stand the theory
remains powerful: it provides a simple yet effective method for correcting insulating and 
semiconducting systems, where DFT exhibits one of its most striking limitations in 
its inability to accurately predict the band gap.

\subsubsection{Empty state localization}
\label{sec:empty_sate_localization}
While the energy functional minimization typically leads to a set of 
very well localized occupied orbitals, this is not the case for the empty states, which,
even at the KC level, turn out to be delocalized\footnote{The empty states resulting
from the KI functionals are delocalized due to (i) the entanglement of the high-lying nearly
free electron bands (which are very delocalized) and low-lying conduction bands, and (ii) 
the residual Hartree contribution to empty states' potentials (see the detailed description
of the KC potentials in Ref.~\citenum{borghi_koopmans-compliant_2014}).}. 
Applying the KC corrections on delocalized
empty states would lead to corrective terms that vanish in the limit of infinite 
systems~\cite{nguyen_koopmans-compliant_2018}, thus leaving the unoccupied band structure
totally uncorrected and identical to the one of the underlying density functional approximation.
Using a localized set of orbitals is indeed a key requirement to deal 
with extended systems, and to get KC corrections to the band structure that remain 
finite (rather than tend to zero) and converge rapidly to their thermodynamic 
limit~\cite{nguyen_koopmans-compliant_2018}. For this reason we typically compute a 
non-self-consistent Koopmans correction using maximally localized Wannier functions
as the localized representation for the lower part of the empty manifold~\cite{de_gennaro_blochs_2021, nguyen_koopmans-compliant_2018}. This heuristic
choice provides a practical and effective scheme, as clearly 
supported by the results of previous works~\cite{nguyen_koopmans-compliant_2018, de_gennaro_blochs_2021} 
and confirmed here. Moreover, it does not affect the occupied 
manifold and therefore does not change the potential energy surface of the functional.

\subsection{Energy functional minimization}
\label{sec:minimization}
The algorithm used to minimize any KC functional consists of two nested steps~\cite{borghi_variational_2015}, 
inspired by the ensemble-DFT approach~\cite{marzari_ensemble_1997}: First,
(i) a minimization is performed with respect to all unitary transformations of the 
orbitals at fixed manifold, i.e. leaving unchanged the Hilbert subspace spanned by these
orbitals (the so-called ``inner-loop''). This minimizes
the orbital-density-dependent contribution to the KC functional. Then (ii) an optimization 
of the orbitals in the direction orthogonal to the subspace is performed via a standard conjugate-gradient algorithm
(the so-called ``outer-loop'').
This two steps are iterated, imposing throughout the orthonormality of the orbitals, until the minimum
is reached. To speed up the convergence, the minimization is typically performed starting
from a reasonable guess for the variational orbitals. As discussed above, for extended 
systems a very good choice for this guess are the MLWFs calculated from the ground state of 
the base functional. For these orbitals the screening coefficients are calculated and 
kept fixed during the minimization. Ideally, these can be recalculated at the end of 
the minimization if the variational orbitals changed significantly, thus implementing
a full self-consistent cycle for the energy minimization. 

While this is the most 
rigorous way to perform a KC calculation, in the next section we will we resort to two well-controlled
approximations to simplify the formalism and make it possible to use an efficient implementation
in primitive cell: i) we use a second order Taylor expansion of Eq.~\ref{eq:KI} and ii)
we assume that the variational orbitals coincide with MLWFs from the underlying density
functional. The first assumption allows us to replace expensive $\Delta$SCF calculations in a
supercell with cheaper primitive cell ones using DFPT, while the second allows us to skip altogether the minimization of the 
functional, while still providing a very good approximation for the variational 
orbitals~\cite{borghi_variational_2015, nguyen_koopmans-compliant_2018, de_gennaro_blochs_2021}.
A formal justification of the second order Taylor expansion is discussed in Sec.~\ref{sec:KmeetW}
and its overall effect on the final results is discussed in Sec.~\ref{sec:validation} and in the
Supporting Information.

\section{A simplified KI implementation: Koopmans meets Wannier}
\label{sec:KmeetW}

In previous work on the application of KC functionals to periodic 
crystals~\cite{nguyen_koopmans-compliant_2018} the calculation of the screening coefficients and the 
minimization of the KC functional were performed using a supercell approach. While this is a very 
general strategy (and the only possible one for non-periodic system), for periodic solids it 
is desirable to work with a primitive cell, exploiting translational symmetry and thus 
reducing the computational cost. The obstacle to this (and the reason for the previous supercell approach) is 
the localized nature of the variational orbitals and the orbital-density-dependence of the KI 
Hamiltonian which apparently breaks the translational symmetry of the crystal. Nevertheless, one can 
argue that the Bloch symmetry is still preserved~\cite{heaton_self-interaction_1983,  almbladh_exact_1985}
which allows the variational orbitals to be expressed as Wannier functions~\cite{wannier_structure_1937} (WFs).
The translational properties of the WFs can then be exploited to recast the supercell problem into a primitive cell one plus
a sampling of the Brillouin zone.

In the present implementation, we use a Taylor expansion of Eq.~\ref{eq:KI} retaining only
the terms up to second order in $f_i$~\cite{colonna_screening_2018, salzner_koopmans_2009,stein_curvature_2012,zhang_orbital_2015, mei_exact_2021}. While this 
approximation is not strictly necessary, it allows us to simplify the expression for
the KI corrections and potentials, and at the same time it does not affect the dominant Hartree 
contribution in Eq.~\ref{eq:KI}, which is exactly quadratic in the
occupations. The residual difference in the xc contribution has a minor effect on
the final results (see section~\ref{sec:validation}). The unrelaxed KI energy corrections
and potentials become~\cite{colonna_screening_2018, colonna_koopmans-compliant_2019}
\begin{align}
    \Pi^{\rm KI(2)}_i & = \frac{1}{2}f_i(1-f_i) \langle n_i | f_{\rm Hxc} | n_i \rangle \label{eq:ene_ki2_u} \\
    \mathcal{V}_i^{\rm KI(2)}(\br) & = \frac{\delta \Pi^{\rm KI(2)}_i}{\delta \rho_i(\br)} = -\frac{1}{2}\langle n_i | f_{\rm Hxc} | n_i \rangle + \nn \\
    & + (1-f_i) \int d\br ' f_{\rm Hxc}(\br, \br') n_i(\br')
    \label{eq:pot_ki2_u}
\end{align}
where the superscript ``$^{\rm (2)}$'' underscores the fact that this is a second-order 
expansion of the full KI energy and potential. 

We note that the DFT kernel $f_{\rm Hxc}$
depends only on the total charge density and therefore has the periodicity of the primitive cell, 
while the variational  orbitals are periodic on the supercell. Based on the translational 
symmetry of perfectly periodic systems the assumption can be made that variational 
orbitals can be expressed as WFs~\cite{de_gennaro_blochs_2021}.
%: $| \phi_n \rangle \rightarrow |\bR n\rangle $.
By definition the WFs $\omega_{\bR n}(\br)$ are labeled
according to the lattice vector $\bR$ of the home cell inside the supercell; have the
periodicity of the supercell, i.e. $\omega_{\bR n}(\br)=\omega_{\bR n}(\br+\bT)$  with
$\bT$ any lattice vector of the supercell; and are such that $\omega_{\bR n}(\br) = \omega_{\bzero n}(\br -\bR)$.
The WFs provides an alternative but completely equivalent description of the
electronic structure of a crystal, via a unitary matrix transformation of the
delocalized Bloch states $\psi_{\mathbf{k}n}$:
\begin{align}
    w_{\mathbf{R}n}(\mathbf{r}) & = \frac{1}{N_\mathbf{k}} \sum_\mathbf{k} e^{-i\mathbf{k}\cdot\mathbf{R}} \psi_{\mathbf{k}n}(\mathbf{r}) \nn \\
     & = \frac{1}{N_\mathbf{k}} \sum_\mathbf{k} e^{-i\mathbf{k}\cdot\mathbf{R}} e^{i\mathbf{k}\cdot\mathbf{r}} w_{\mathbf{k}n}(\mathbf{r}) \nn \\
     w_{\mathbf{k}n}(\mathbf{r}) &= \sum_v U^{(\bk)}_{nm} u_{\mathbf{k}m}(\mathbf{r}). 
     \label{eq:MLWF_def}
\end{align}
In this expression $w_{\bk n}(\br)= \sum_{v} U^{(\bk)}_{nm} u_{\bk m}(\br)$ is a very general
``gauge transformation'' of the periodic part of the canonical Bloch state $u_{\bk m}(\br)$, 
$N_{\bk}$ is the number of $\bk$ points and $\bR$ the Bravais lattice vectors of the primitive cell. The 
expression above highlights the duality between variational orbitals (Wannier functions) and 
canonical orbitals (Bloch states), and the simple connection between the two. In periodic 
systems the transformation relating these two sets of orbitals can be decomposed in a phase 
factor $e^{- i \bk \cdot \bR}$ and a $\bk$-dependent unitary rotation mixing only Bloch states
at the same $\bk$. This unitary matrix is defined in principle by the minimization of the 
orbital-density-dependent correction to the energy functionals (see
Sec.~\ref{sec:minimization}). However, this minimization greatly increases the computational cost of
these calculations relative to functionals of the electronic density alone.
As discussed in 
section~\ref{sec:basic_features}, the minimization of the KI functional (in the limit of an 
infinitesimally small PZ-SIC term) leads to localized orbitals that closely resemble MLWFs~\cite{borghi_variational_2015}. 
For this reason we make a further assumption and assume that the unitary matrix defining 
the variational orbitals can be obtained via a standard Wannierization procedure, i.e. 
by minimizing the sum of the quadratic spread of the Wannier functions~\cite{marzari_maximally_1997, marzari_maximally_2012},
thus allowing us to bypass the computationally intense energy minimization.
Under this assumption the KI functional closely resembles related
approaches like the Wannier-Koopmans~\cite{ma_using_2016} and the
Wannier-transition-state methods~\cite{anisimov_transition_2005}. In both these schemes
the linearity of the energy is enforced when adding/removing an electron from a set of
Wannier functions, resulting in accurate prediction of band gaps and band structure of
a variety of systems~\cite{anisimov_transition_2005, ma_using_2016, weng_wannier_2017, weng_wannier_2018, li_wannier-koopmans_2018, weng_wannierkoopmans_2020}.
At variance with these two methods, the present approach is based on a variational 
expression of the total energy (Eq.~(\ref{eq:KC_lin})) as a function of the orbital
densities which automatically leads to a set of Wannier-like variational orbitals.
Moreover, from a practical point of view, within the present implementation the 
evaluation of the energy and potential corrections can be efficiently evaluated 
using density functional perturbation theory, as detailed in the next section, 
thus avoiding expensive supercell calculations typically needed for both
the Wannier-Koopmans~\cite{ma_using_2016} and the Wannier-transition-state methods~\cite{anisimov_transition_2005}. 
 
Overall, in this simplified framework, all the ingredients are then provided by a standard
DFT calculation followed by a Wannierization of the canonical KS-DFT eigenstates. The KI 
calculation reduces then to a one-shot procedure where the screening coefficients in Eq.~\ref{eq:alpha_lr} and the KI Hamiltonian specified by Eq.~\ref{eq:KC_ham}
and Eq.~\ref{eq:pot_ki2_u} needs to be evaluated on the localized representation provided by the MLWFs. 
This can be done straightforwardly by working in a supercell to accommodate the 
real-space representation of the Wannier orbital densities, or, as pursued here, by working in reciprocal
space and exclusively within the primitive cell, thus avoiding expensive supercell calculations. This latter
strategy leverages the translational properties of the Wannier functions. By expressing the Wannier orbital
densities as Bloch sum in the primitive cell as described in Sec.~\ref{sec:rho_wann}, we must then
i) recast the equation for the screening coefficients (cf. Eq.~\ref{eq:alpha_lr}) into a linear
response problem suitable for an efficient implementation 
 using the reciprocal space formulation of density-functional perturbation theory as detailed
 in Sec.~\ref{sec:screen_coeff}, and ii) devise, compute and diagonalize the KI Hamiltonian
 at each $\bk$-point in the BZ of the primitive cell as illustrated in Sec.~\ref{sec:ki_ham}.

%%%%%%%%%%%%%%%%%%%%%%%%%%%%%%
\subsection{From Wannier orbitals in the supercell to Bloch sums in the primitive cell}
\label{sec:rho_wann}
%%%%%%%%%%%%%%%%%%%%%%%%%%%%%%
%
\begin{figure*}[t]
 \begin{center}
    \includegraphics[width=0.9\textwidth]{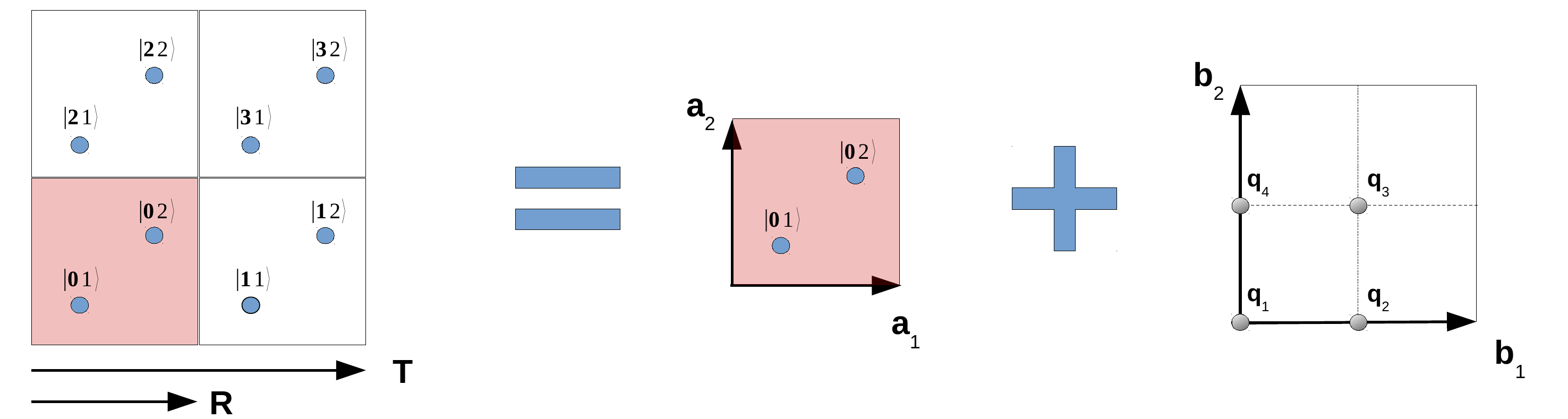}
    \caption{Schematic representation of the supercell --- primitive cell mapping in 2D. A $2 \times 2$ 
    supercell problem with a $\Gamma$ only sampling can be recast in a primitive cell problem with a 
    $2 \times 2$ sampling of the Brillouin zone.}
    \label{fig:sc-pc}
 \end{center}
\end{figure*}

The first step in this reformulation is to rewrite the Wannier orbital densities as Bloch sums in 
the primitive cell. A schematic view of the supercell-primitive cell mapping is shown in Fig.~\ref{fig:sc-pc}. Using the definition of the MLWFs, the Wannier orbital densities can be written as
\begin{align}
    \rho_{\bR n}(\br) & = |\omega_{\bR n}(\br)|^2  = \left| \frac{1}{N_{\bk}}\sum_{\bk} e^{-i \bk \cdot \bR} \tilde{\psi}_{\bk n}(\br)\right|^2 \nonumber \\
                            & = \frac{1}{N_{\bq}}\sum_{\bq} e^{i \bq \cdot \br} \left\{ e^{-i \bq \cdot \bR} \frac{1}{N_{\bk}}\sum_{\bk} w^*_{\bk, n}(\br)w_{\bk+\bq, n}(\br) \right\} \nonumber \\
                            & = \frac{1}{N_{\bq}}\sum_{\bq} e^{i \bq \cdot \br} \rho_{\bq}^{\bR n}(\br).
    \label{eq:wann_orb_dens}
\end{align}
Since the $w_{\bk n}(\br)$ are periodic on the primitive cell, the $\rho^{\bR n}_{\bq}(\br)$ are also, and
consequently the Wannier orbital density is given as a sum over the Brillouin zone (BZ) of 
primitive cell-periodic function just modulated by a phase factor $e^{i \bq \cdot \br}$. The periodic
densities $\rho^{\bR n}_{\bq}(\br) = \rho^{\bR n}_{\bq}(\br+\bR)$ are the basic ingredients
needed to express integrals over the supercell appearing in the definitions of the screening 
coefficients and of the KI corrections and potentials into integrals over the primitive cell. 

%%%%%%%%%%%%%%%%%%%%%%%%%%%%%%%%%%%%%%
\subsubsection{Screening coefficients}
\label{sec:screen_coeff}
%%%%%%%%%%%%%%%%%%%%%%%%%%%%%%%%%%%%%%
The expression for the screening coefficients given in Eq.~(\ref{eq:alpha_lr}) can be 
recast in a linear response-problem~\cite{colonna_screening_2018} suitable for an 
efficient implementation based on DFPT~\cite{baroni_phonons_2001}: 

\begin{align}\label{eq:alpha}
    \alpha_{\bzero n} & = \frac{ \langle \rho_{\bzero n} | \left[ \epsilon^{-1}f_{\rm Hxc} \right] | \rho_{\bzero n} \rangle}{\langle \rho_{\bzero n} | \left[ f_{\rm Hxc} \right] | \rho_{\bzero n} \rangle} 
    = 1 + \frac{\langle V^{\bzero n}_{\rm pert} | \Delta^{\bzero n} \rho \rangle}{\langle \rho_{\bzero n} | V^{\bzero n}_{\rm pert} \rangle}. 
\end{align}
In the expression above we made use of the definition of the dielectric matrix 
$\epsilon^{-1} = 1 + f_{\rm Hxc}\chi$ with $\chi$ being the density-density response 
function of the system at the underlying DFT level; $\Delta^{\bzero n} \rho(\br) = 
\int d\br' \chi(\br, \br') V_{\rm pert}^{\bzero n}(\br')$ is by definition the density
response induced in the systems due to the ``perturbing potential'' $V^{\bzero n}_{\rm pert}(\br) =
\int d\br' f_{\rm Hxc} (\br, \br') \rho_n(\br')$. This perturbation represents the 
Hartree-exchange-correlation potential generated when an infinitesimal fraction of 
an electron is added to/removed from a MLWF. The perturbing potential has the same 
periodic structure as the Wannier density in Eq.~(\ref{eq:wann_orb_dens}) and can 
be decomposed into a sum of monochromatic perturbations in the primitive cell, 
$V_{\rm pert}^{\bzero n}(\br) = \sum_{\bq} e^{i\bq \cdot \br} V^{\bzero n}_{{\rm pert},\bq}(\br)$ with 
\begin{equation}
 V^{\bzero n}_{{\rm pert}, \bq}(\br) = \int d\br' f_{\rm Hxc}(\br, \br') \rho^{\bzero n}_{\bq}(\br') .
 \label{eq:Vpert_KC}
\end{equation}
The total density variation $\Delta^{\bzero n} \rho(\br')$ induced by the bare perturbation
$V_{\rm pert}^{\bzero n}(\br)$ reads
\begin{align}
    \Delta^{\bzero n} \rho(\br) & = \int d\br' \chi(\br, \br') V_{\rm pert}^{\bzero n}(\br') \nn \\
     & =  \int d\br' \chi(\br, \br') \sum_{\bq} e^{i\bq \cdot \br'}  V_{{\rm pert},\bq}^{\bzero n}(\br') \nn \\
     & = \sum_{\bq} e^{i\bq \cdot \br} \Delta_{\bq}^{\bzero n}\rho(\br)
    \label{eq:dens_var}
\end{align}
where we used the fact that for a periodic system $\chi$ can be decomposed in a sum of primitive cell-periodic
functions $\chi(\br ,\br') = \sum_{\bq} e^{i\bq \cdot (\br - \br')}\chi_{\bq}(\br ,\br')$ ~\footnote{
More in detail: 
\begin{align*}
    &\int d\br' \chi(\br, \br') \sum_{\bq} e^{i\bq \cdot \br'}  V_{{\rm pert},\bq}^{\bzero n}(\br') = \\
    = &\sum_{\bq} \int d\br' e^{i\bq \cdot \br} e^{-i\bq \cdot \br}\chi(\br, \br')  e^{i\bq \cdot \br'}  V_{{\rm pert},\bq}^{\bzero n}(\br') = \\
   = &\sum_{\bq} e^{i\bq \cdot \br} \int d\br' \chi_{\bq}(\br, \br') V_{{\rm pert},\bq}^{\bzero n}(\br') = \\
   = & \sum_{\bq} e^{i\bq \cdot \br} \Delta_{\bq}^{\bzero n}\rho(\br)
\end{align*}
}. 
The primitive cell-periodic density variation is given by:
\begin{align}
    \Delta_{\bq}^{\bzero n} \rho(\br) & = \int d\br' \chi_{\bq}(\br,\br') V_{{\rm pert},\bq}^{\bzero n}(\br') \nn \\ 
    &= \sum_{\bk v} \psi_{\bk,v}^*(\br) \Delta \psi_{\bk+\bq, v}(\br) + c.c.
    \label{eq:dens_var_per}
\end{align}
where $ \Delta\psi_{\bk+\bq}(\br)$ is the first order variation of the KS orbitals due to the
perturbation (the bare one plus the SCF response in the Hxc potential). Only the projection 
of the variation of the KS wave functions on the conduction manifold contributes to the 
density response in Eq.~(\ref{eq:dens_var_per}), meaning that $\Delta\psi_{\bk+\bq}(\br)$ 
can be thought of as its own projection on this manifold and it is given by the solution
of the following linear problem~\cite{baroni_phonons_2001}:
\begin{align}
    & \left( H +\gamma P_v^{\bk+\bq} -\varepsilon_{\bk,v} \right) \Delta \psi_{\bk+\bq, v}(\br) \nn \\
     = & -P_c^{\bk+\bq} \left[V_{{\rm pert},\bq}^{\bzero n}(\br) + \Delta_{\bq} V_{\rm Hxc}(\br)\right]\psi_{\bk,v}(\br)
    \label{eq:lin_eq}
\end{align}
where $H$ is the ground state KS Hamiltonian, $P_v^{\bk}=\sum_v |u_{\bk, v}\rangle \langle u_{\bk, v}|$ 
and $P_c^{\bk} = 1-P_v^{\bk}$ are the projectors onto the occupied- and empty-manifold respectively, 
$\gamma$ is a constant chosen in such a way that the $\gamma P_v^{\bk+\bq}$ operator makes the 
linear system non-singular~\cite{baroni_phonons_2001}, and 
\begin{equation}
    \Delta_{\bq} V_{\rm Hxc}(\br) = \int d \br ' f_{\rm Hxc}(\br, \br') \Delta^{\bzero n}_{\bq}\rho (\br')
    \label{eq:Delta_Vscf}
\end{equation} 
is the self-consistent variation of the Hxc potential due to the charge 
density variation $\Delta_{\bq}^{\bzero n}\rho$. Iterating 
Eqs.~(\ref{eq:dens_var_per})-(\ref{eq:Delta_Vscf}) to self-consistency leads 
to the final results for $ \Delta\rho_{\bq}^{\bzero n}(\br)$ and the screening
coefficient is finally obtained by summing over all the $\bq$ contributions:
\begin{align}
    \alpha_{\bzero n} =  1 + \frac{\sum_{\bq} \langle V^{\bzero n}_{{\rm pert},\bq} |  \Delta^{\bzero n}_{\bq}\rho \rangle} {\sum_{\bq} \langle {\rho^{\bzero n}_{\bq}} | V^{\bzero n}_{{\rm pert}, \bq} \rangle} .
    \label{eq:alpha_gspace}
\end{align}
Equations 
~(\ref{eq:Vpert_KC})-(\ref{eq:Delta_Vscf}) show how to recast the calculation 
of the screening coefficient into a linear response problem in the primitive cell that can
be efficiently solved using the machinery of DFPT ,and are key to the present work.
Linear-response equations for different $\bq$ are decoupled and the original 
problem is thus decomposed into a set of independent problems that can be 
solved on separate computational resources, allowing for straightforward 
parallelization. More importantly, the computational cost is also greatly 
reduced: In a standard supercell implementation the screening coefficients are computed with a 
finite difference approach by performing additional total-energy calculations 
where the occupation of a Wannier function is 
constrained~\cite{nguyen_koopmans-compliant_2018}. This requires, for each 
MLWF, multiple SCF calculations with a computational time $T^{\rm SC}$ that 
scales roughly as ${N_{\rm el}^{\rm SC}}^3$, where $N_{\rm el}^{\rm SC}$ is 
the number of electrons in the supercell. The primitive cell DFPT approach described above scales
instead as $T^{\rm PC} \propto N_{\bq} N_{\bk} {N_{\rm el}^{\rm PC}}^3$; this 
is the typical computational time for the SCF cycle ($N_{\bk} {N_{\rm el}^{\rm PC}}^3$), 
times the number of independent monochromatic perturbations 
($N_{\bq}$). Using the relation $N_{\rm el}^{\rm SC}=N_{\bk}N_{\rm el}^{\rm 
PC}$, and the fact that $N_{\bq}=N_{\bk}$, the ratio between the supercell and primitive cell 
computational times is $T^{\rm SC}/T^{\rm PC} \propto N_{\bq}$. Therefore as 
the supercell size (and, equivalently, the number of $\bq$-points in the primitive cell) 
increases, the primitive cell DFPT approach becomes more and more computationally 
convenient.
We finally point out that a similar strategy was recently implemented in the 
context of the linear-response approach to the calculation of the Hubbard 
parameters in DFT+U~\cite{cococcioni_linear_2005} in order to avoid the 
use of a supercell~\cite{timrov_hubbard_2018, timrov_self-consistent_2021}.

%%%%%%%%%%%%%%%%%%%%%%%%%%%%%%%%%%%%%%%%%%%%%%%%%%%%%%%%%%%%%%%%%%%%%%%%%%%%%%%%%%%%%%%%%%%%%%%%%%%%%%
%\begin{figure*}[t]
%    \centering
%    \includegraphics[width=0.32\textwidth]{Figs/more/Self_Hxc.png}
%    \includegraphics[width=0.32\textwidth]{Figs/more/Self_scrHxc.png}
%    \includegraphics[width=0.32\textwidth]{Figs/more/alpha.png}
%    \caption{Effectiveness of the Gygi-Baldereschi scheme for the evaluation of the bare and screened %Hartree integrals in the Li $s$-like Wannier function of the LiF FCC crystal. The numbers below the red %circles indicates the number of $\bk$-point in
%    the regular mesh along each direction.}
%    \label{fig:GB_LiF}
%\end{figure*}
%%%%%%%%%%%%%%%%%%%%%%%%%%%%%%%%%%%%%%%%%%%%%%%%%%%%%%%%%%%%%%%%%%%%%%%%%%%%%%%%%%%%%%%%%%%%%%%%%%%%%%

%%%%%%%%%%%%%%%%%%%%%%%%%%%%%%
\subsubsection{KI Hamiltonian}
\label{sec:ki_ham}
%%%%%%%%%%%%%%%%%%%%%%%%%%%%%%
As it is typical for orbital-density-dependent functionals, the canonical eigenvalues and 
eigenvectors are given by the diagonalization of the matrix of Lagrangian multipliers 
$H_{mn}(\bR)=\me{\bR m}{\hat{h}_{\bzero n}}{\bzero n}$ with $\hat{h}_{\bzero m}|\bzero n \rangle = 
\delta E^{\rm KI(2)}/\delta \langle \bR n|$. In the case of insulating systems, the 
matrix elements between occupied and empty states vanish~\cite{dabo_piecewise_2014} 
and we can treat the two manifolds separately. For occupied states, the KI potential
is simply a scalar correction (the second term in Eq.~(\ref{eq:pot_ki2_u}) is 
identically zero if $f_i=1$), and thus the KI contribution to the Hamiltonian is diagonal and $\bR$-independent:
\begin{align}
    \Delta H^{\rm KI(2)}_{vv'}(\bR) & = \me{\bR v}{\mathcal{V}^{\rm KI(2)}_{\bzero v'}}{\bzero v'} \nn \\
    %& = -\frac{1}{2} \langle \rho_{\bzero v} | \left[ f_{\rm Hxc} \right] | \rho_{\bzero v} \rangle \delta_{\bR,\bzero} \delta_{vv'} \nn \\
    & =-\frac{1}{2N_{\bq}} \sum_{\bq} \langle {\rho^{\bzero v}_{\bq}} | V^{\bzero v}_{{\rm pert}, \bq} \rangle \delta_{\bR,\bzero} \delta_{vv'} \nn \\
    & = -\frac{1}{2}\Delta^{\rm KI(2)}_{\bzero v} \delta_{\bR,\bzero} \delta_{vv'} 
\end{align}
Using the definition of $| \bR v \rangle$ or equivalently the identity $\delta_{\bR, \bzero} = 
1/N_{\bk}\sum_{\bk} e^{i\bk \cdot \bR}$ in the equation above, the KI contribution to the
Hamiltonian at a given $\bk$ can be identified as: 
\begin{align}
    \Delta H^{\rm KI(2)}_{vv'}(\bk) & = -\frac{1}{2}\Delta^{\rm KI(2)}_{\bzero v} \delta_{vv'} %\nonumber \\
    %& =  -\frac{1}{2} \delta_{vv'} \sum_{\bq} \langle {\rho^{\bzero v}_{\bq}} | V^{\bzero v}_{{\rm pert}, \bq} \rangle 
    \label{eq:ki-ham-K-occ}
\end{align}
which is $\bk$-independent, as expected.

In the case of empty states, in addition to the scalar term in the equation above, there is also a  
non-scalar contribution~\cite{borghi_koopmans-compliant_2014} that needs a more careful
analysis. This term is given by the matrix element of the non-scalar contribution to the KI 
potential, i.e. $\mathcal{V}^{\rm KI(2),r}_{\bzero c}(\br) = \int d\br' f_{\rm Hxc}(\br, \br') \rho_{\bzero c}(\br')$, and reads: 
\begin{align}
    \Delta H^{\rm KI(2),r}_{cc'}(\bR) 
    & = \me{\bR c}{\mathcal{V}^{\rm KI(2), r}_{\bzero c'}}{\bzero c'} \nn \\
    & = \frac{1}{N_{\bk}}\sum_{\bk} e^{i\bk \cdot \bR} \left[ \frac{1}{N_{\bq}}\sum_{\bq} \langle V^{\bzero c'}_{{\rm pert}, \bq}| \rho^{cc'}_{\bk, \bk+\bq}\rangle \right]
    \label{eq:ki-ham-R-emp-real}
\end{align}
where $\rho^{cc'}_{\bk, \bk+\bq}(\br) = w^*_{\bk,c}(\br)w_{\bk+\bq,c'}(\br)$ (see Supporting 
Information for a detailed derivation of the KI matrix elements). Since the dependence on the 
$\bR$-vector only appears in the complex exponential, the matrix elements of the KI Hamiltonian
in $\bk$-space can be easily identified as the term inside the square brackets in 
Eq.\eqref{eq:ki-ham-R-emp-real}. Including the scalar contribution leads to the 
$\bk$-space Hamiltonian for the empty manifold: 
\begin{equation}
    \Delta H^{\rm KI(2)}_{cc'}(\bk) =-\frac{1}{2}\Delta^{\rm KI(2)}_{\bzero c} \delta_{cc'} + \frac{1}{N_{\bq}}\sum_{\bq} \langle V^{\bzero c'}_{{\rm pert}, \bq}| \rho^{cc'}_{\bk, \bk+\bq}\rangle
    \label{eq:ki-ham-K-emp}
\end{equation}
Eqs.~(\ref{eq:ki-ham-K-occ}) and~(\ref{eq:ki-ham-K-emp}) define the KI contribution to the 
Hamiltonian at a given $\bk$ point on the regular mesh used for the sampling of the Brillouin zone.
This contribution needs to be scaled by the screening coefficient $\alpha_{\bzero n}$ and added to  
the DFT Hamiltonian to define the full KI Hamiltonian at $\bk$ as:
\begin{equation}
    H^{\rm KI(2)}_{mn}(\bk) = H^{\rm DFT}_{mn}(\bk) + \alpha_{\bzero n} \Delta H^{\rm KI(2)}_{mn}(\bk)
    \label{eq:ki-ham-2nd}
\end{equation}
where $H^{\rm DFT}_{mn}(\bk) = \me{w_{\bk m}}{\hat{H}^{\rm DFT}}{w_{\bk n}}$ is the KS-DFT
Hamiltonian evaluated on the periodic part of the Bloch states in the Wannier gauge 
(see Eq.~(\ref{eq:MLWF_def})). The diagonalization of $H^{\rm KI(2)}_{mn}(\bk)$ defines
the canonical KI eigenstates $\{\psi_{\bk i}^{\rm KI(2)}; \varepsilon_{\bk i}^{\rm KI(2)}\}$. 
Finally, given the localized nature of the MLWFs it is also possible to interpolate the 
Hamiltonian with standard techniques~\cite{slater_simplified_1954,souza_maximally_2001,marzari_maximally_2012}
to obtain the KI eigenvalues at any arbitrary $\bk$ point in the Brillouin zone.

\subsection{Technical aspects of the implementation}
The calculations of the screening parameters and KI potentials involve the evaluation of 
bare and screened Hxc integrals of the form $\me{\rho_{\bzero n}}{f_{\rm Hxc}}{\rho_{\bzero n}}$
and $\me{\rho_{\bzero n}}{\epsilon^{-1}f_{\rm Hxc}}{\rho_{\bzero n}}$. Because of the 
long-range nature of the Hartree kernel, these integrals are diverging in periodic-boundary
conditions (PBC) and therefore require particular caution. The divergence can be avoided 
adding a neutralizing background (in practice this means that the $\bq+\bG =\bzero$ 
component of the Hartree kernel is always set to zero). The integrals are then finite 
and, more importantly, converge to the correct electrostatic energy of isolated Wannier
functions.~\cite{makov_periodic_1995} However, the convergence is extremely slow ($1/N_{\bk}^{1/3}$
to leading order) because of the $1/|\bq+\bG|^2$ divergence in the Coulomb kernel. This
is a well know problem and many solutions have been proposed to overcome it; e.g. Makov
and Payne~\cite{makov_periodic_1995} (MP) suggested to remove from the energy the
electrostatic interaction of a periodically-repeated point-charge. Other approaches
consist in truncating the Coulomb interaction~\cite{ismail-beigi_truncation_2006, rozzi_exact_2006}
or using the scheme proposed by Martyna and Tuckermann~\cite{martyna_reciprocal_1999} or the charge
or density corrections of Ref.~\citenum{dabo_electrostatics_2008}. Here we adopt the approach
devised by Gygi and Baldereschi (GB)~\cite{gygi_quasiparticle_1989} that consists of adding 
and subtracting to the integrand a function with the same divergence and whose integral can
be computed analytically. The result is a smooth function suitable for numerical integration,
plus an analytical contribution. From a computational point of view this amounts to defining
a modified Hartree kernel 
\begin{equation}
f_{\rm H}(\bq+\bG) = 
  \begin{cases}
    D, & \text{if } \bq+\bG = \bzero \\
    4\pi/|\bq+\bG|^2, & \text{if } \bq+\bG \neq \bzero 
  \end{cases}    
\end{equation}
where $D$ is the reciprocal-space part of the Ewald sum for a point charge, repeated according
to the super-periodicity defined by the grid of $\bq$-points~\cite{nguyen_efficient_2009}.
For the screened Hartree integral the $\bq+\bG=0$ component needs to be further scaled by the 
macroscopic dielectric function $\epsilon_{\infty}$ of the system~\footnote{This is only strictly
valid for cubic systems, where the dielectric tensor is diagonal with $\epsilon^{(11)}_{\infty} =
\epsilon^{(22)}_{\infty} = \epsilon^{(33)}_{\infty}$. In the most general case a generalization
of the Ewald technique must be used~\cite{fischerauer_comments_1997, rurali_theory_2009, murphy_anisotropic_2013} }
because in this case we are dealing with the screened Coulomb integral $\epsilon^{-1}f_H$. In this work
we compute $\epsilon_{\infty}$ from first-principles~\cite{baroni_phonons_2001} using the PHONON code of \QE{}.

Another important point is that the periodic part of the Wannier function at 
$\bk$ and $\bk + \bq$ must come from the same Wannierization procedure, 
otherwise the localization property of the Wannier orbital density 
[Eq.~(\ref{eq:wann_orb_dens})] will be lost because of unphysical phase 
factors possibly due to the diagonalization routine or other computational 
reasons. This requirement is enforced using a uniform grid centered at 
$\Gamma$ such that $\bk+\bq = \bp + \bG$ with $\bp$ still belonging to the 
original grid and $\bG$ a reciprocal lattice vector. In this way 
$w_{\bk+\bq}(\br)$ can be obtained from $w_{\bp}(\br)$ simply multiplying it 
by a phase factor $e^{-i\bG \cdot \br}$. As a direct consequence of this 
choice the mesh of $\bq$ points for the LR calculation has to be the same as 
the one used for the Wannierization.

Finally, in order to be compliant with the current limitation of working with 
block-diagonal occupation matrices (see. Sec.~\ref{sec:limitation_metals}), 
the Wannierization procedure needs to be prevented from mixing the occupied 
and empty manifolds, so that the occupation matrix retains its block-diagonal 
form in the localized-orbital representation. In practice this is done by 
performing two separate Wannierizations, one for the occupied and one for the 
empty manifold. 
To obtain the maximally localized Wannier orbitals for the low-lying empty states,
we employed the disentanglement maximally localized Wannier function technique 
proposed in Ref.~\citenum{souza_maximally_2001}.

%%%%%%%%%%%%%%%%%%%%%%%%%%%%%%%%%%%%%%%%%%%%%%%%%%%%%%%%%%%%%%%%%%%%%%%%%%%%%%%%%%%%%%%%%%%%%
\begin{table}[hb]
 \centering
 \begin{tabularx}{\linewidth}{*{6}{>{\centering\arraybackslash}X}}% <-- changed
    \hline
    \hline
    Sys. & wann & method & $\frac{\partial^2 E}{\partial f^2}[eV]$ & $\frac{d^2 E}{df^2}[eV]$ & $\alpha$  \\
    \hline
    Si & $sp^3$ (V)    & FD & 3.906 & 0.888 & 0.227  \\
       &               & DFPT & 3.886 & 0.887 & 0.228  \vspace{0.1cm} \\
       & $sp^3$ (C)    & FD & 1.351 & 0.215 & 0.160  \\
       &               & DFPT & 1.351 & 0.218 & 0.162  \vspace{0.1cm} \\
    \hline 
    GaAs & $d$ (V)      & FD & 10.159 & 3.530 & 0.347  \\
         &              & DFPT & 10.217 & 3.550 & 0.347  \vspace{0.1cm} \\
         & $sp^3$ (V)   & FD & 3.899 & 0.976 & 0.250  \\
         &              & DFPT & 3.896 & 0.936 & 0.240 \vspace{0.1cm} \\
         & $sp^3$ (C)   & FD & 1.418 & 0.233 & 0.164  \\
         &              & DFPT & 1.372 & 0.243 & 0.177  \\
    \hline
 \end{tabularx}
 \caption{Comparison between analytical (computed with DFPT) and numerical [computed with finite differences (FD)]
 second derivatives of the energy with respect to the occupation of different kinds of Wannier functions for 
 silicon (Si) and gallium arsenide (GaAs). V / C indicates whether the MLWF is from the valence or conduction
 manifold. Partial $\partial/\partial f_i$ and full $d/df_i$ derivatives refer to unrelaxed and relaxed 
 quantity, respectively.}
 \label{tab:alpha}
\end{table}
%%%%%%%%%%%%%%%%%%%%%%%%%%%%%%%%%%%%%%%%%%%%%%%%%%%%%%%%%%%%%%%%%%%%%%%%%%%%%%%%%%%%%%%%%%%%%%%

\section{Results and discussion}
In this section we first validate the present implementation against a standard KI one as 
described in Ref.~\citenum{nguyen_koopmans-compliant_2018}, and then discuss the application
to few paradigmatic test cases, highlighting the advantages and limitations of the approach. 
In particular we analyze the band structure of gallium arsenide (GaAs), hexagonal wurtzite
zinc oxide (ZnO) and face-centered-cubic (FCC) lithium fluoride (LiF) at three levels of 
theory: i) the local density approxination (LDA), ii) the hybrid-functional scheme by 
Heyd Scuseria and Ernzerhof (HSE)~\cite{heyd_hybrid_2003, heyd_erratum:_2006}, and 
iii) the KI functional within the implementation described in this work. All calculations
are performed using the plane-wave (PW) and pseudopotential (PP) method as implemented
in the \QE{} package~\cite{giannozzi_quantum_2009, giannozzi_advanced_2017}. The LDA 
functional is used as the underlying density-functinal approximation for all the KI 
calculations. LDA scalar relativistic Optimized Norm-conserving Vanderbilt PPs~\cite{hamann_optimized_2013,hamann_erratum_2017}
from the DOJO library~\cite{van_setten_pseudodojo_2018} are used to model the interaction
between the valence electrons and the nucleus plus the core electrons~\footnote{The LDA
pseudopotentials are available at \href{http://www.pseudo-dojo.org/}{www.pseudo-dojo.org}.
Version 0.4.1., standard accuracy}. 
Maximally localized Wannier functions are computed using the Wannier90 code~\cite{pizzi_wannier90_2020}.
For all the systems we used the experimental crystal structures taken from the inorganic Crystal Structure
Database~\footnote{ICSD website, \href{http://www.fiz-karlsruhe.com/icsd.html}{http://www.fiz-karlsruhe.com/icsd.html}} 
(ICSD); GaAs, ZnO and LiF correspond to ICSD numbers 107946, 162843 and 62361 respectively.
For the LDA calculations we used a $10 \times 10 \times 10$ $\bk$-point mesh for GaAs, a 
$12 \times 12 \times 7$  $\bk$-point mesh for ZnO and a $14 \times 14 \times 14$ $\bk$-point mesh 
for LiF. The kinetic energy cutoff for the PW expansion of the wave-functions is set to 
$E_{\rm cut} = 80$ Ry (320 Ry for the density and potentials expansion) in all cases. 
For the HSE calculations we verified that a reduced cutoff $E^{\rm Fock}_{\rm cut}= 120$ 
and a $\bk$-point grid typically twice as coarse as the LDA one are sufficient for the 
convergence of the exchange energy and potential. For the screening parameters and KI 
Hamiltonian calculations we used the same energy cutoff and a $\bq$-mesh of $6 \times 6 \times 6$
for GaAs and LiF, and a $6 \times 6 \times 4$ mesh for ZnO.

%%%%%%%%%%%%%%%%%%%%%%%%%%%%%%%%%%%%%%%%%%%%%%%%%%%%%%%%%%%%%%%%%%%%%%%%%%%%%%%%%%%%%%%%%%%%%%%
\begin{figure}[t]
    \centering
    \includegraphics[width=0.45\textwidth]{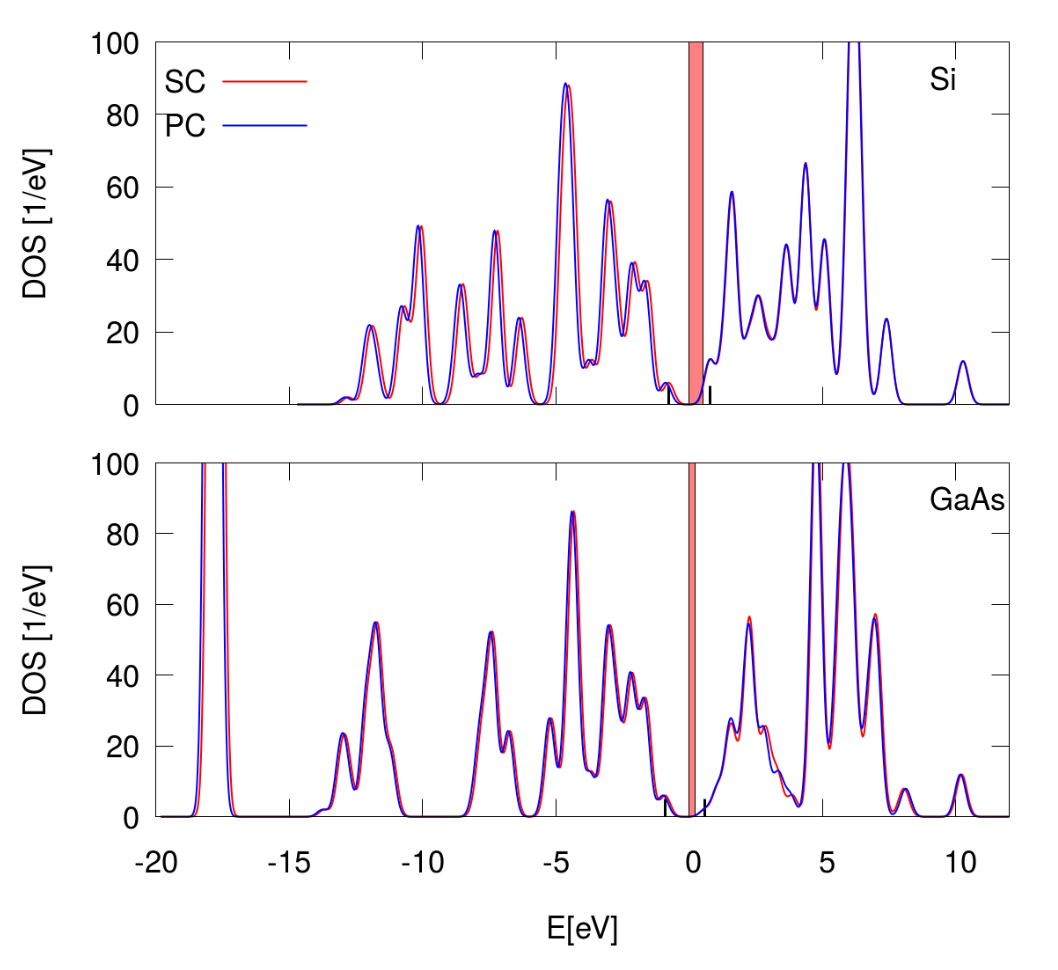}
    \caption{KI density of states of Si (upper panel) and GaAs (lower panel) computed with a reference 4$\times$4$\times$4 supercell (red line) and with the present approach working in the primitive cell with a commensurate 4$\times$4$\times$4 $\bk$-points mesh.}
    \label{fig:dos}
\end{figure}
%%%%%%%%%%%%%%%%%%%%%%%%%%%%%%%%%%%%%%%%%%%%%%%%%%%%%%%%%%%%%%%%%%%%%%%%%%%%%%%%%%%%%%%%%%%%%%%

%%%%%%%%%%%%%%%%%% GaAs Band Structure %%%%%%%%%%%%%
 \begin{figure*}[t]
 \textbf{GaAs band structure}\par\medskip
    \begin{subfigure}{}
        \includegraphics[width=0.3\textwidth]{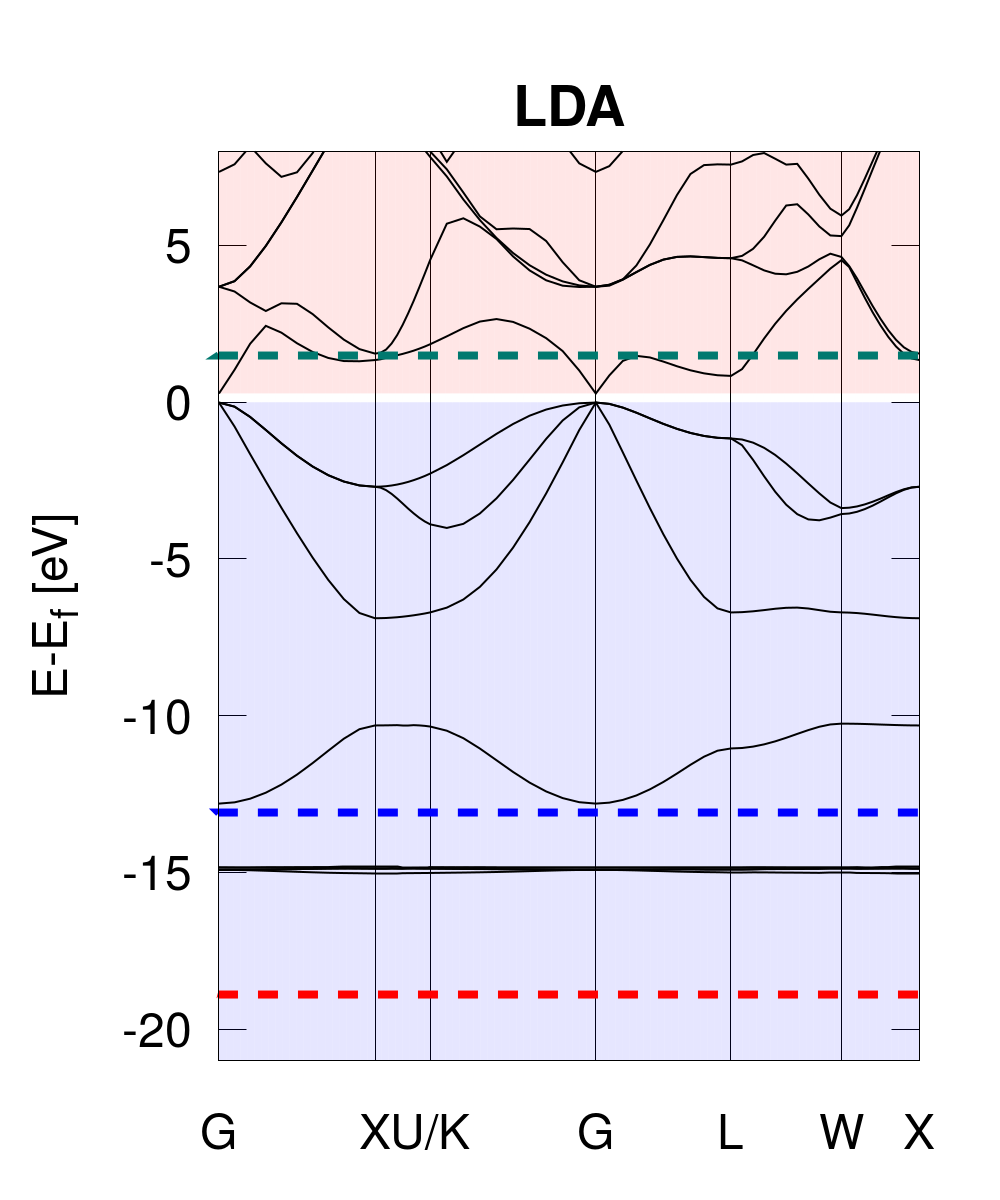}
        \includegraphics[width=0.3\textwidth]{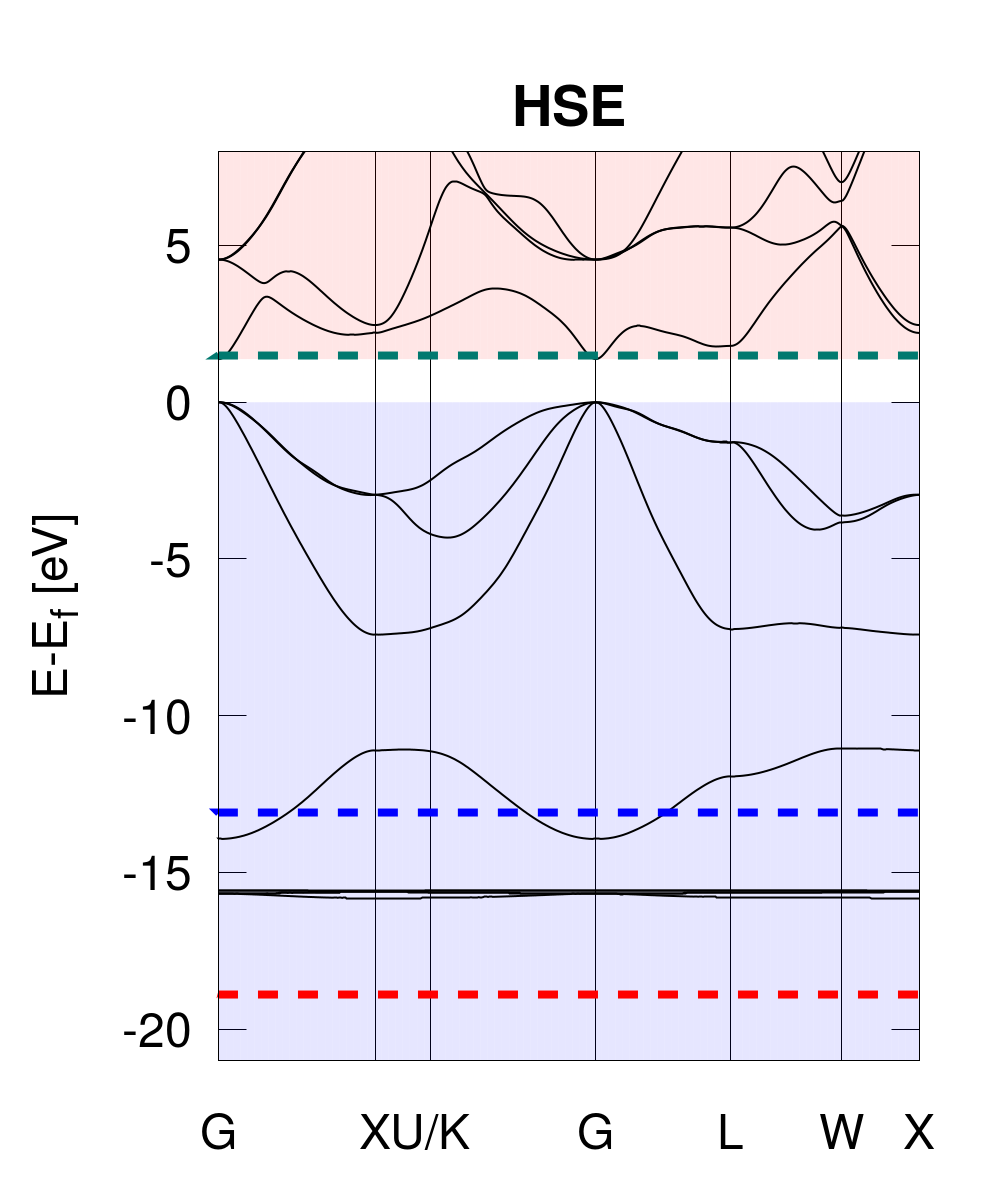}
        \includegraphics[width=0.3\textwidth]{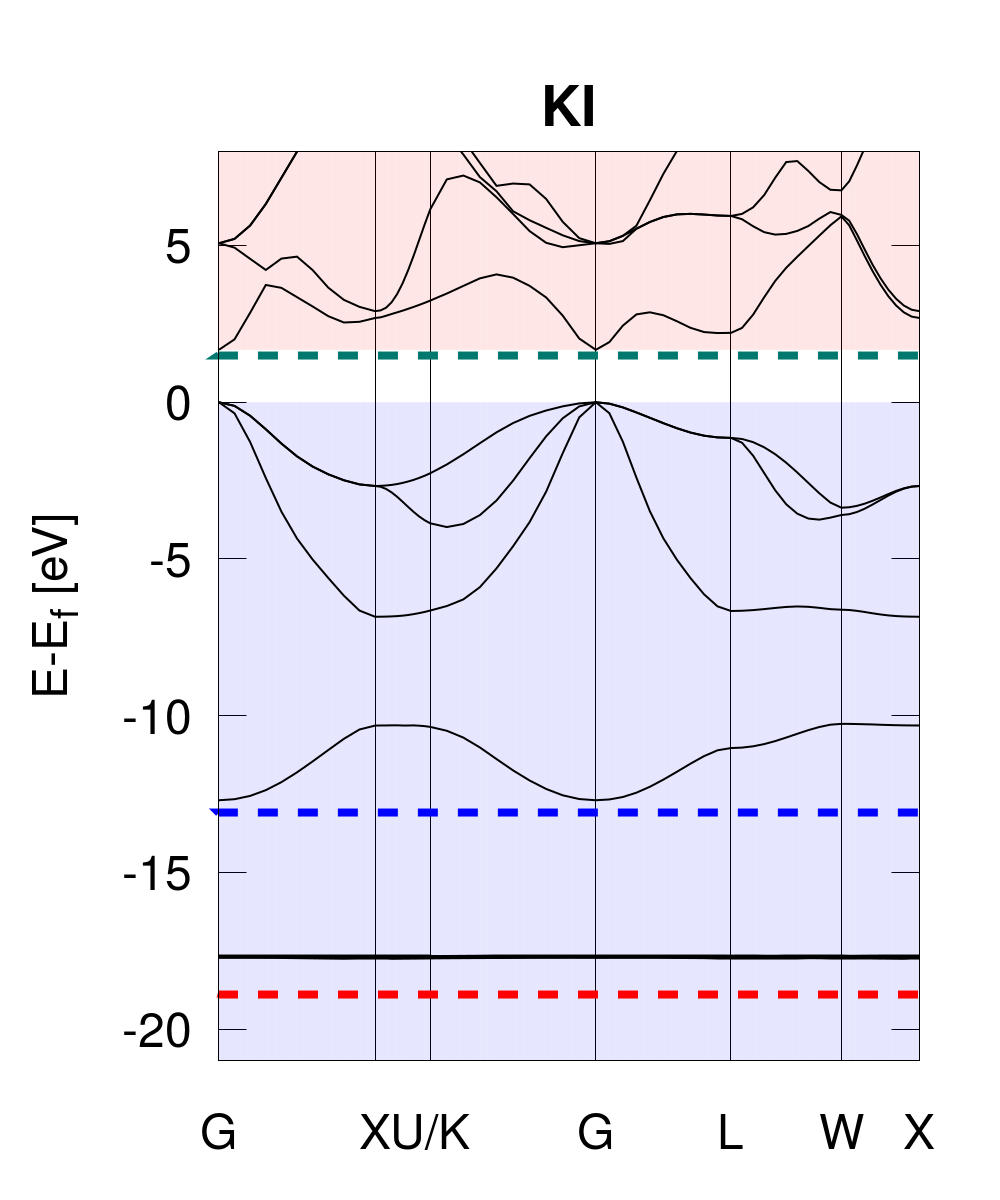}
    \end{subfigure}
    \begin{subfigure}{} 
        \renewcommand\tabularxcolumn[1]{m{#1}}% <-- added
        \renewcommand\arraystretch{1.3}
        \setlength\tabcolsep{2pt}% <-- added
    \begin{tabularx}{\linewidth}{*{7}{>{\centering\arraybackslash}X}}% <-- changed
        \hline
        \hline
        & LDA & HSE & GW$_0$ & scG$\tilde{\rm W}$ & KI & Exp. \\
        \hline
        E$_{\rm gap}$(eV) & 0.19 & 1.28 & 1.55  & 1.62 & 1.57 & 1.52 \\
        $\langle \varepsilon_d \rangle$(eV) & -14.9 & -15.6 & -17.3 & -17.6 & -17.7 & -18.9 \\
        $W$(eV) & 12.8 & 13.9 & -- & -- & 12.8 & 13.1 \\
        \hline
    \end{tabularx}
    \end{subfigure}
\caption{Band structure of GaAs calculated at different levels of theory:
LDA (left panel), HSE (middle panel) and KI (right panel). Shaded areas
highlight valence (light blue) and conduction (light red) manifolds. The
experimental values for the band gap, valence band width, and energy position of 
Ga $d$-states are represented by the dashed green, blue and red lines, respectively.
Table: Band gap, position of Ga $d$ states with respect to the top of the valence band,
and valence band width ($W$) at different level of theory compared to 
experimental~\cite{shevchik_densities_1974, madelung_semiconductors_2004} and GW results
from Ref.~\citenum{shishkin_accurate_2007}. Theoretical values of the band gap are
corrected for spin-orbit coupling (0.10 eV).}
\label{fig:ki_gaas_bands}
\end{figure*}
%%%%%%%%%%%%%%%%%%%%%%%%%%%%%%%%%%%%%%%%%%%%%%%%%%%%%%%%
\subsection{Validation}
\label{sec:validation}
In order to validate the implementation of the analytical formula for the derivatives
based on the DFTP [Eq.~(\ref{eq:alpha})], we compare the result with a finite difference
calculation where we add/remove a tiny fraction of an electron from a given Wannier 
function. This is done using a  4$\times$4$\times$4 supercell, consistent with the $\bk$/$\bq$
mesh in the primitive cell calculation. In Table~(\ref{tab:alpha}) we present the results for two 
semiconductors, silicon (Si) and gallium arsenide (GaAs). For Si the Wannierization 
produces four identical bonding $sp^3$-like MLWFs spanning the occupied manifold and
four anti-bonding $sp^3$-like MLWFs spanning a four-dimensional manifold disentangled
from the lowest part of the conduction bands. In the case of GaAs we obtained 5 $d$-like
and 4 $sp^3$-like MLWFs representing the occupied manifold and 4 anti-bonding $sp^3$-like
MLWFs from the lowest part of the empty manifold. The numerical and analytical values for
the derivatives agree within few hundredths of an eV. The residual discrepancy is possibly
due to tiny differences in the Wannierization procedure (for the supercell a specific algorithm 
for a $\Gamma$-only calculation was used), and to the difficulties in converging to arbitrary
accuracy the constrained calculations in the supercell.

In order to quantify the error introduced by the second-order approximation adopted here, we 
compare in Fig.~(\ref{fig:dos}) the KI density of states for Si and GaAs computed using
a 4$\times$4$\times$4 supercell within the original implementation~\cite{nguyen_koopmans-compliant_2018, de_gennaro_blochs_2021},
and the present approach working in primitive cell with a consistent 4$\times$4$\times$4 $\bk/\bq$-points mesh. 
For this figure the single particle eigenvalues were convoluted with a Gaussian function
with a broadening of 0.2 eV. The zero of the energy is set to the LDA valence band maximum (VBM),
and the shaded red area represent the LDA band gap. The thick black ticks on the energy axes
mark the position of the KI VBM and conduction band minimum (CBM). The KI VBM and CBM are 
shifted downwards and upwards with respect to the corresponding LDA quantities, leading 
to an opening the fundamental band gap, that goes from 0.51 eV to 1.41 eV and from 0.20
eV to 1.57 eV for Si and GaAs, respectively. We stress here that these results are not 
fully converged with respect to the BZ sampling (or supercell size), and serve just as a 
validation test. The two DOS are in very good agreement, but small differences between
the reference supercell and the primitive cell calculations are present. In particular there is a small 
downward shift of the order or 0.05 eV in the valence part of the DOS, and also tiny 
differences in the conduction one, especially evident for GaAs. These discrepancies 
are due to the second order approximation used for the calculation of the screening 
parameters and KI Hamiltonian in the primitive cell implementation (additional details are provided
in Supporting Information). Nevertheless, all the main features of the DOS are correctly
reproduced, thus validating the present implementation.

\subsection{Application to selected systems}
\noindent \textit{\bf Gallium arsenide:} GaAs is a III-V direct band gap semiconductor
with a zincblende crystal structure. The band structure around the band gap is dominated
by $s$ and $p$ orbitals from Ga and As forming $sp^3$ hybrid orbitals while the flat
bands around 18.9 eV below the VBM are from the $d$ states of Ga. All these features
are correctly reproduced by the LDA band structure [see Fig.~(\ref{fig:ki_gaas_bands})] 
but the band gap $E^{\rm LDA}_{\rm gap} =0.28$ eV is too small, the average $d$ states
position $\langle \varepsilon^{\rm LDA}_d \rangle= -14.9$ eV is too high, and the 
valence bad width $W^{\rm LDA}=12.8$ eV is slightly underestimated. The HSE functional
corrects these errors to some extent, opening the band gap up to $E^{\rm HSE}_{\rm gap}
=1.38$ eV, and shifting downwards the Ga $d$ states, $\langle \varepsilon^{\rm HSE}_d \rangle=-15.6$ eV,
but it also over-stretches the valence band thus overestimating the valence band
width ($W^{\rm HSE} =13.9$ eV). These discrepancies with respect to experiment 
are possibly due to the fact that the fraction of Fock exchange and the range-separation
parameter defining any hybrid scheme might have to be in principle system- and possibly
state-dependent~\cite{skone_self-consistent_2014,brawand_generalization_2016,brawand_performance_2017}.
On the contrary the parameters of the HSE functionals (and also of the vast majority of hybrid schemes)
are system-independent and this is probably not sufficient for an accurate description of all the
spectral features mentioned above. 
The KI functional with its orbital-dependent corrections produces an upward shift of the conduction
manifold and a state-dependent downward shift of the valence manifold (with respect to the LDA VBM),
leading to a better agreement with experimental data for $E_{\rm gap}$ and $\langle \varepsilon_d \rangle$.
The $sp^3$ band width is instead identical to that of the underlying density functional
approximation (LDA), which is already in good agreement with the experimental value. 
This is because of the scalar nature of the KI corrections for the occupied manifold 
combined with the fact that the Wannierization produces four identical but differently-oriented
$sp^3$ MLWFs spanning the four uppermost valence bands. Then from Eq.~(\ref{eq:ki-ham-K-occ})
it follows that the KI contribution to the Hamiltonian is not only $\bk$-independent but
also the same for all the $sp^3$ bands. The KIPZ functional~\cite{borghi_koopmans-compliant_2014},
another flavor of KC functionals, might correct this because of its non-scalar contribution
to the effective potentials. This will introduce an off-diagonal coupling between different
bands and will thus modify the band width~\cite{de_gennaro_blochs_2021}.
Overall the KI results are in extremely good agreement with experimental data. This
performance is even more remarkable if compared to GW results~\cite{shishkin_accurate_2007}
reported in the Table under Fig.~\ref{fig:ki_gaas_bands}, with KI showing the same accuracy
as self-consistent GW plus vertex correction in the screened Coulomb interaction.

%%%%%%%%%%%%%%%%%% LiF Band Structure %%%%%%%%%%%%%
 \begin{figure*}[t]
    \textbf{LiF band structure}\par\medskip
    \begin{subfigure}{}
        \includegraphics[width=0.3\textwidth]{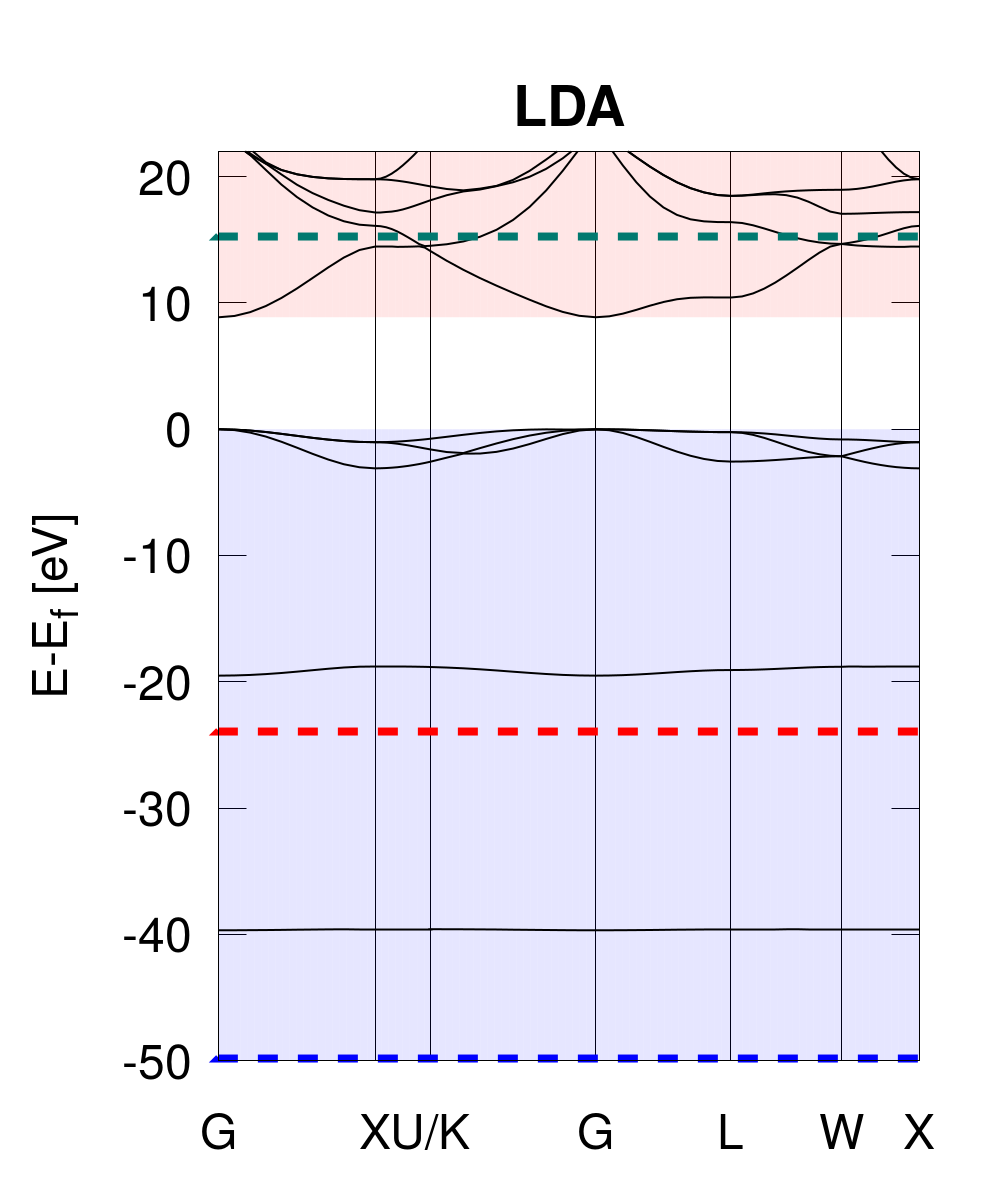}
        \includegraphics[width=0.3\textwidth]{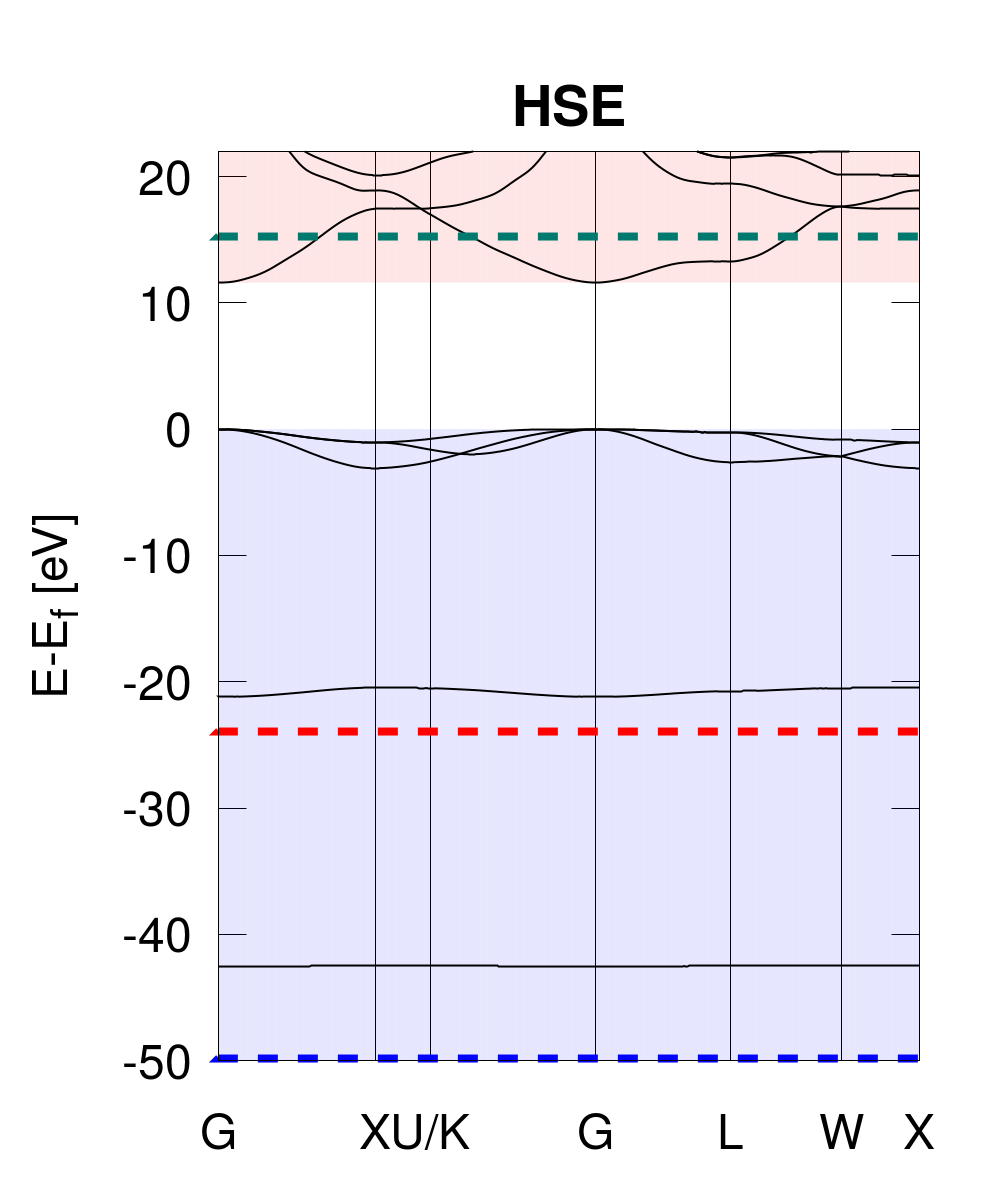}
        \includegraphics[width=0.3\textwidth]{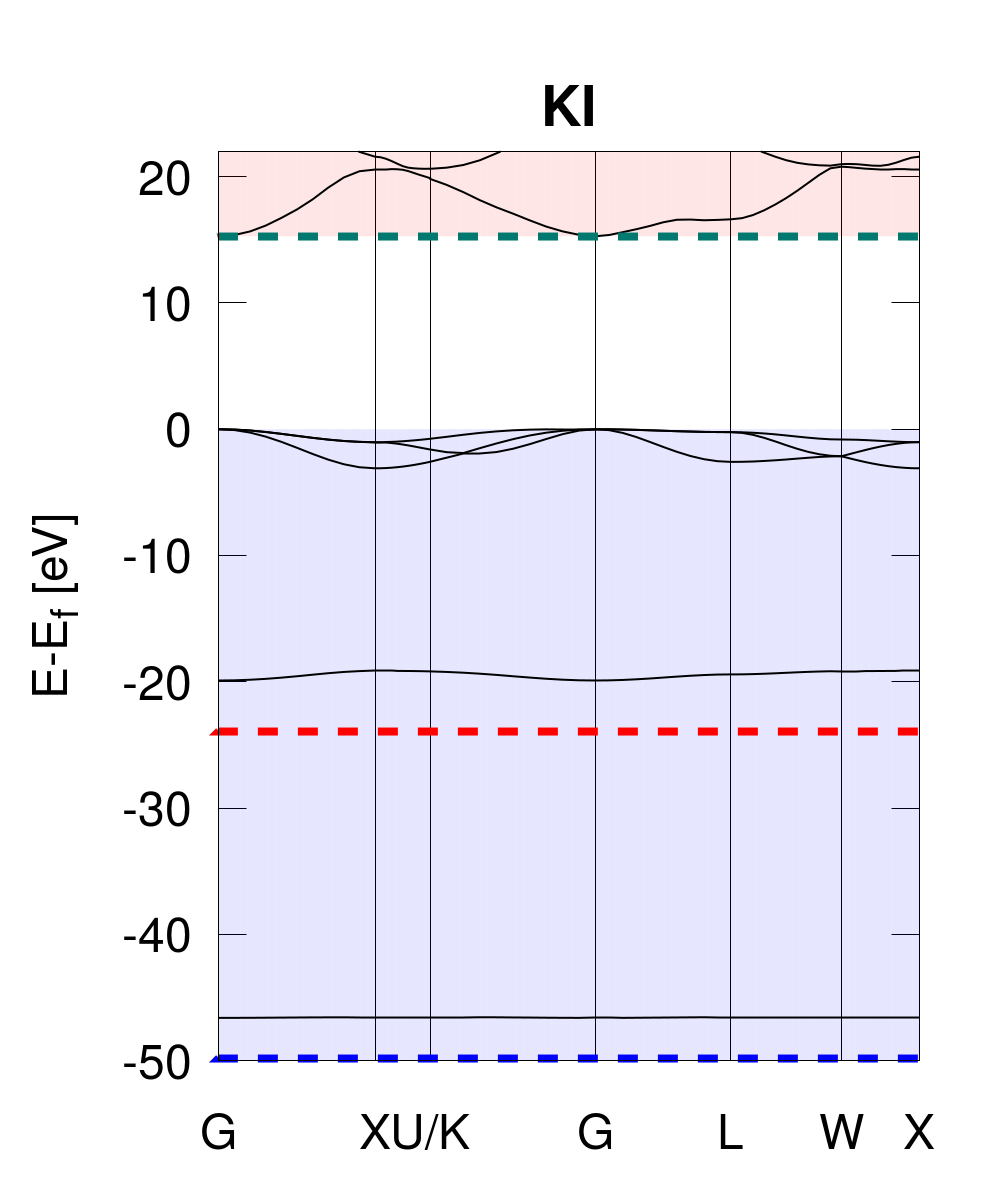}
    \end{subfigure}
    \begin{subfigure}{} 
        \renewcommand\tabularxcolumn[1]{m{#1}}
        \renewcommand\arraystretch{1.3}
        \setlength\tabcolsep{2pt}
    \begin{tabularx}{\linewidth}{*{7}{>{\centering\arraybackslash}X}}
        \hline
        \hline
                          & LDA   & HSE    & GW$_0$ & scG$\tilde{\rm W}$ &    KI   & (Exp.) \\
        \hline
        E$_{\rm gap}$(eV) & 8.87  & 11.61  & 13.96  &  14.5              &  15.28  & 15.35$^{(*)}$\\
        $\langle \varepsilon \rangle_{\rm{F}_{2s}}$(eV) & -19.06 & -20.7 & -24.8$^{(\dagger)}$ &  -- &  -19.5 & -23.9 \\
        $\langle \varepsilon \rangle_{\rm{Li}_{1s}}$(eV) & -39.6 & -42.5 & -47.2$^{(\dagger)}$ &  -- &  -46.6 & -49.8 \\
        \hline
    \end{tabularx}

    \end{subfigure}
\caption{Band structure of LiF calculated at different levels of theory:
LDA (left panel), HSE (middle panel) and KI (right panel). Shaded areas
highlight valence (light blue) and conduction (light red) manifolds. The
experimental value for the band gap, the F $2s$ band and Li $1s$ band are represented
by the dashed green, red, and blue lines, respectively.
Table: Band gap and low lying energy levels at different levels of theory compared to
GW results~\cite{shishkin_accurate_2007, shishkin_self-consistent_2007, wang_quasiparticle_2003}
and to experiments~\cite{piacentini_thermoreflectance_1976, johansson_core_1976}. $^{(*)}$ 
The zero-point-renormalization~\cite{nery_quasiparticles_2018} (-1.15 eV) has been subtracted
to the experimental gap~\cite{piacentini_thermoreflectance_1976} (14.2 eV) to have a meaningful
comparison with the calculations. $(^{\dagger})$ Values obtained at G$_0$W$_0$ level.~\cite{wang_quasiparticle_2003}} 
\label{fig:ki_LiF_bands}
\end{figure*}
%%%%%%%%%%%%%%%%%%%%%%%%%%%%%%%%%%%%%%%%%%%%%%%%%%%%%%%%%%%%%%%%%%%%5

\noindent\textit{\bf Lithium fluoride:} LiF crystallizes in a rock-salt structure and provides a
prototypical example of wide gap insulators with a marked ionic character. Its band structure at
all the different levels of theory analyzed in this work is presented in Fig.~(\ref{fig:ki_LiF_bands}).
LiF is a direct band gap material with the topmost valence bands exclusively contributed by F $2p$
orbitals, and the lower part of the conduction manifolds mainly from Li $2s$ orbitals with a small
contribution from F $2p$ orbitals. The low lying energy levels at about -24 eV and -50 eV with 
respect to the top of the valence bands can be unambiguously classified as F $2s$ and Li $1s$ 
bands, respectively. Also in this case we observe all the limitations of LDA already observed 
and discussed for GaAs and the same trend going from local to hybrid to orbital-density 
dependent KI functionals. In particular, the KI band gap is in very good agreement with 
the experimental band gap~\cite{piacentini_thermoreflectance_1976} after the zero point 
renormalization energy~\cite{nery_quasiparticles_2018} is added to have a fair comparison
between calculations (no electron-phonon effects are accounted for) and experiments. 
Thanks to the state-dependent potentials the Li $1s$ band is pushed downwards in energy
more than the valence bands are, and its relative position with respect to the VBM results
in a much better agreement with experimental values. On the other hand the F$_{2s}$ band
is shifted downwards in energy by roughly the same amount as the three valence bands 
(originating from the F$_{2p}$ states) are. This leaves almost unchanged its distance 
with respect to the VBM (~19.5 eV at KI level compared to -19.06 at LDA level). This 
is at odd with GW results~\cite{wang_quasiparticle_2003} which show an increase in the
relative distance of roughly 5 eV and place the F$_{2s}$ band at -24.8 eV with respect
to the VBM, in better agreement with experimental results~\cite{johansson_core_1976}(-23.9 eV). 
Full KI and KIPZ band structures show the same underestimation although less severe 
(see Supporting Information), especially when using a better underlying density functional
(PBE vs LDA). This seems to suggest this underestimation to be a common feature of the KC
functionals (rather than to an effect solely due to the second order approximation adopted
here) which deserves further investigation. 
Nevertheless the improved description of the band structure close to the Fermi level and 
in particular of the band gap is remarkable also in this case, and a comparison with 
available GW calculations~\cite{shishkin_accurate_2007, shishkin_self-consistent_2007} 
reveals the effectiveness of the KI functional approach presented here as a valid 
alternative to Green's function based methods.

%%%%%%%%%%%%%%%%%%% Si Band Structure %%%%%%%%%%%%%
% \begin{figure*}[t]
%%    \centering
%    \begin{subfigure}{}
%        \includegraphics[width=0.3\textwidth]{Figs/more/Si_lda.png}
%        \includegraphics[width=0.3\textwidth]{Figs/more/Si_hse.png}
%        \includegraphics[width=0.3\textwidth]{Figs/more/Si_ki.png}
%%    \caption{Silicon}
%    \end{subfigure}
%%    \bigskip% <-- added
%    \begin{subfigure}{} %<-- changed width
%%        \centering
%        \renewcommand\tabularxcolumn[1]{m{#1}}% <-- added
%        \renewcommand\arraystretch{1.3}
%        \setlength\tabcolsep{2pt}% <-- added
%    \begin{tabularx}{\linewidth}{*{7}{>{\centering\arraybackslash}X}}% <-- changed
%        \hline
%        \hline
%        & LDA & HSE & GW$_0$ & scG$\tilde{\rm W}$ & KI & Exp. \\
%        \hline
%        E$_{\rm gap}$(eV) & 0.44 & 1.20 & XXX & XXX & 1.21 & 1.17 \\
%        \hline
%    \end{tabularx}
%%        \caption{table}
%    \end{subfigure}
%\caption{Band structure of Si calculated at different level of theory:
%LDA (left panel), HSE (middle panel) and KI (right panel). Shaded areas
%highlight valence (light blue) and conduction (light red) manifolds. The
%experimental values for the band gap is represented by the dashed green line.}
%\label{fig:ki_si_bands}
%\end{figure*}
%%%%%%%%%%%%%%%%%%%%%%%%%%%%%%%%%%%%%%%%%%%%%%%%%%%%%%%%%%%%%%%%%%%%%5

\begin{figure*}[t]
 \textbf{ZnO band structure}\par\medskip
%    \centering
    \begin{subfigure}{}
        \includegraphics[width=0.3\textwidth]{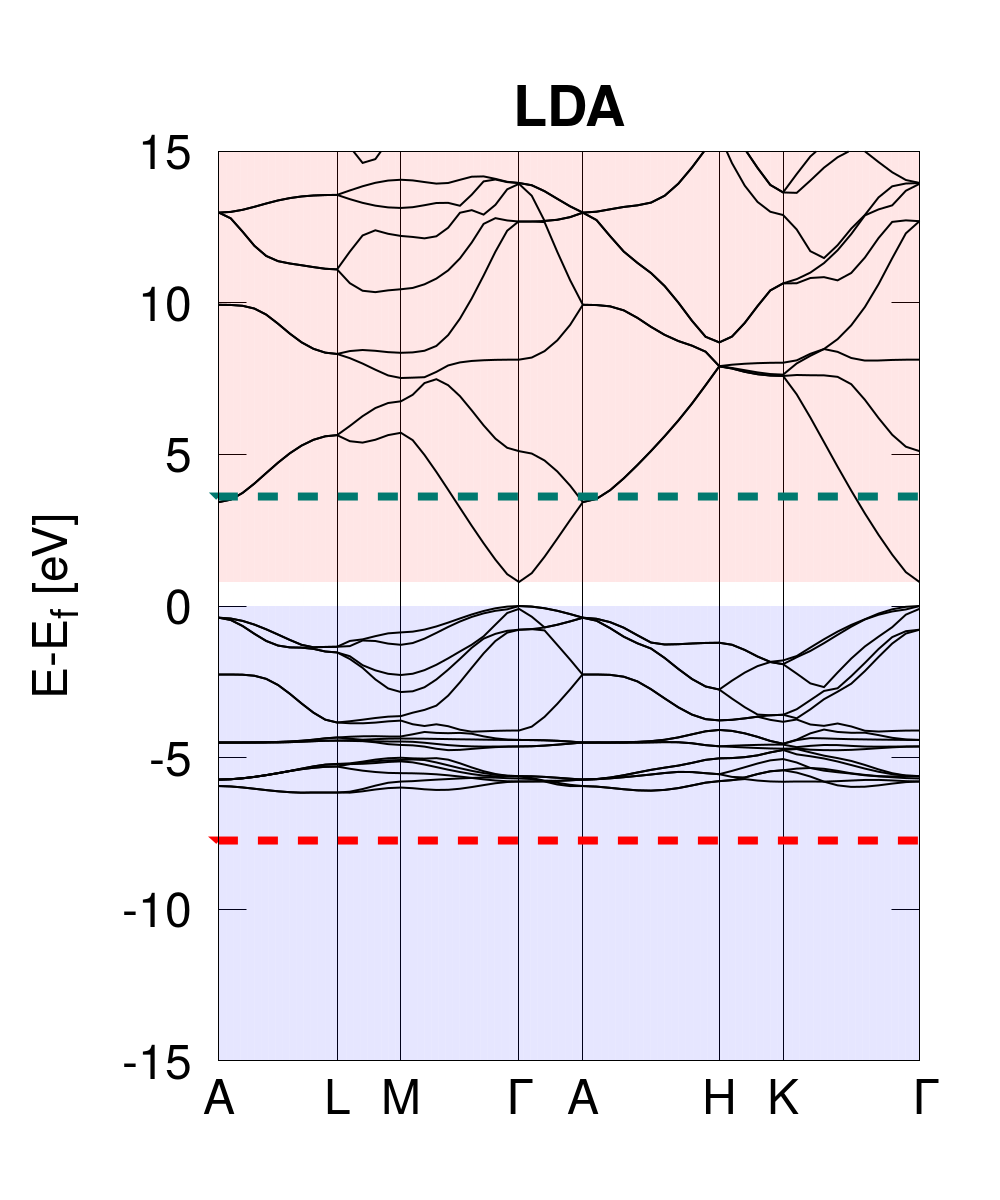}
        \includegraphics[width=0.3\textwidth]{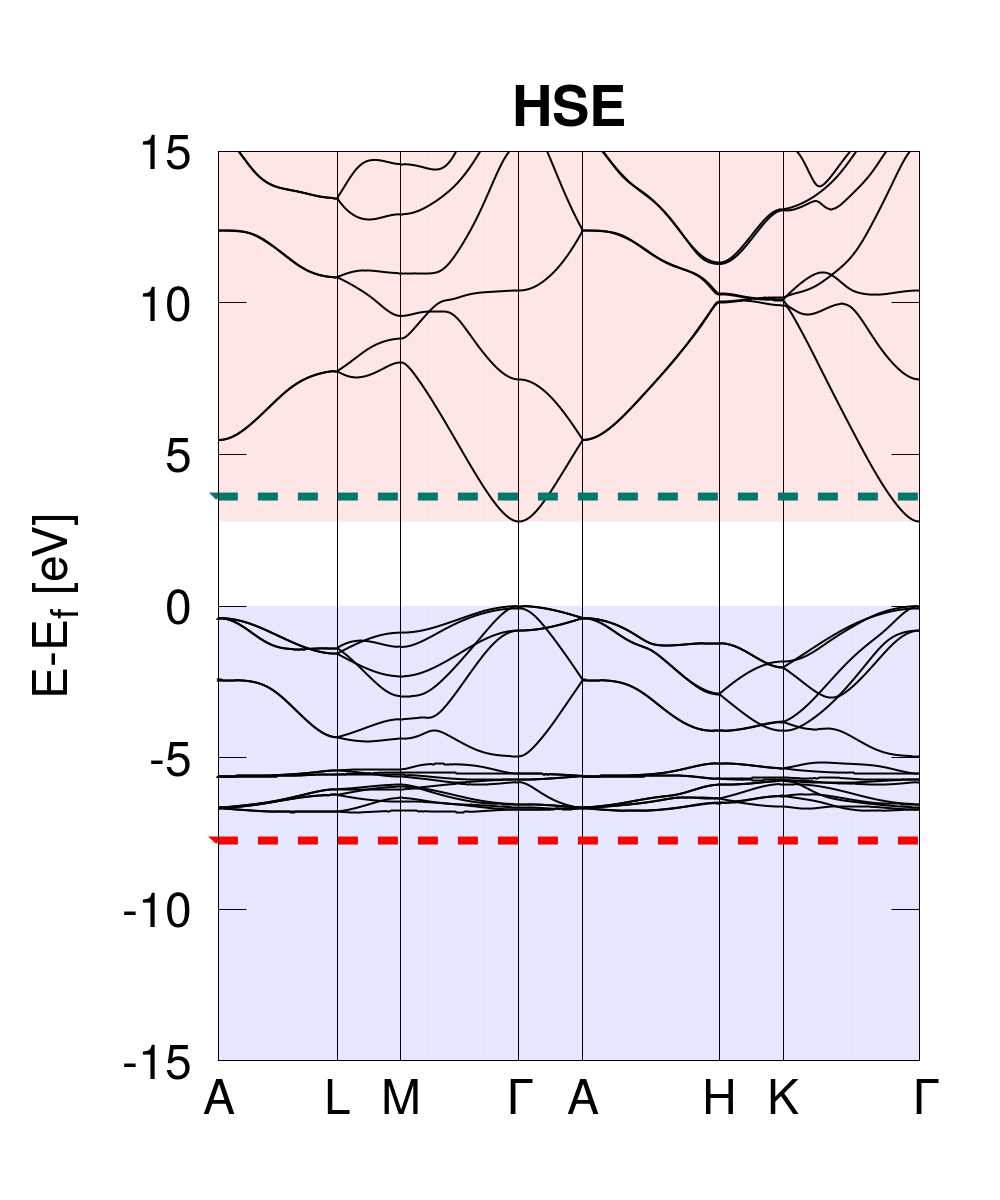}
        \includegraphics[width=0.3\textwidth]{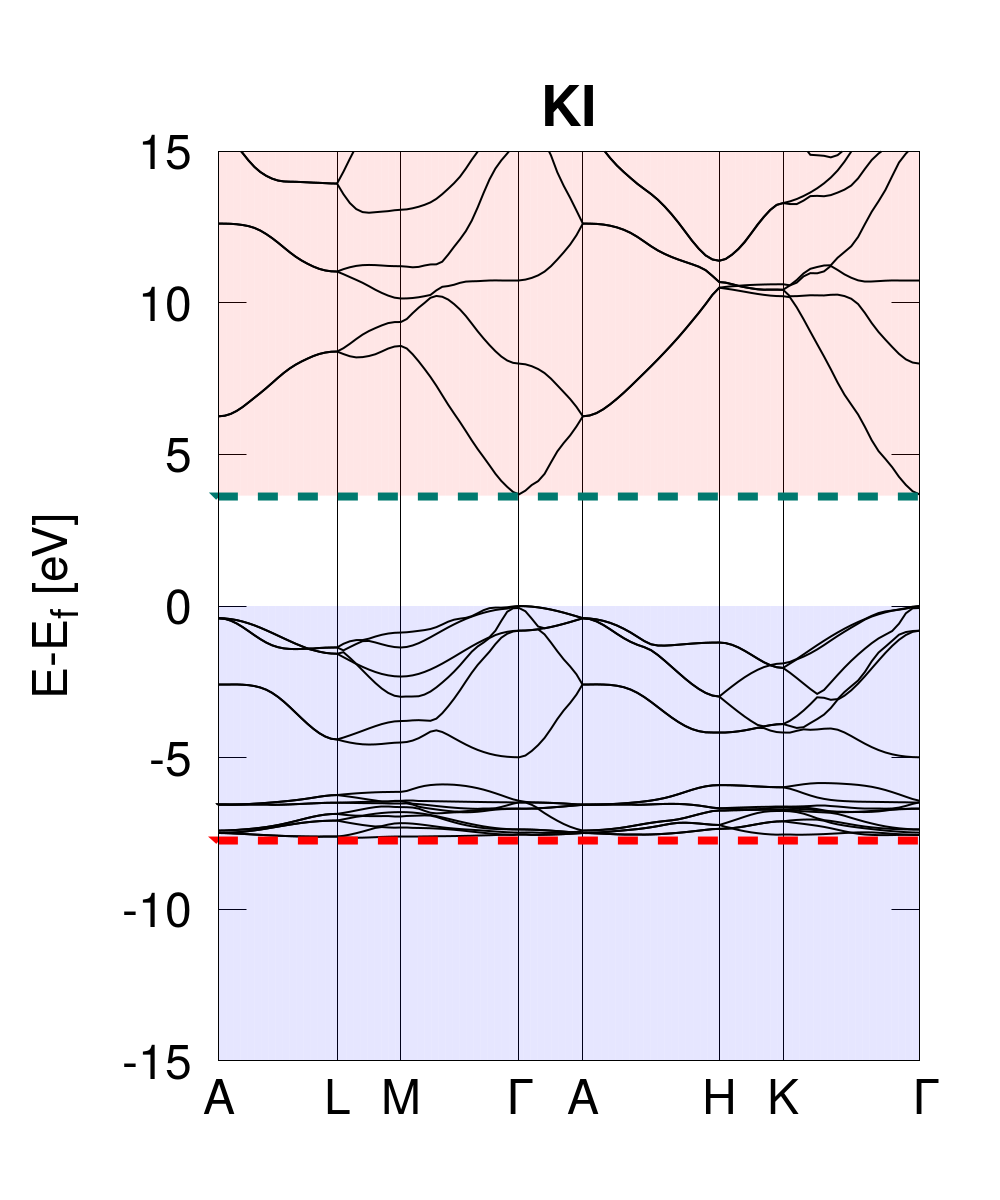}
%    \caption{Silicon}
    \end{subfigure}
%    \bigskip% <-- added
    \begin{subfigure}{} %<-- changed width
%        \centering
        \renewcommand\tabularxcolumn[1]{m{#1}}% <-- added
        \renewcommand\arraystretch{1.3}
        \setlength\tabcolsep{2pt}% <-- added
    \begin{tabularx}{\linewidth}{*{7}{>{\centering\arraybackslash}X}}% <-- changed
        \hline
        \hline
        & LDA & HSE & GW$_0$ & scG$\tilde{\rm W}$ & KI & Exp. \\
        \hline
        E$_{\rm gap}$(eV) & 0.79 & 2.79 & 3.0 & 3.2 & 3.62 & 3.60$^{(*)}$ \\
%\hline
        $\langle \varepsilon_d \rangle$(eV) & -5.1 & -6.1 & -6.4 & -6.7 & -6.9 & -7.5/-8.0 \\
        \hline
    \end{tabularx}
%        \caption{table}
    \end{subfigure}
\caption{Band structure of ZnO calculated at different level of theory:
LDA (left panel), HSE (middle panel) and KI (right panel). Shaded areas
highlight valence (light blue) and conduction (light red) manifolds. The
experimental values for the band gap and for the energy position of 
Zn $d$-states are represented by the dashed green line and by the dashed
red line, respectively.
Table: Band gap and position of Zn $d$ states with respect to the top of the valence 
band at different level of theory compared to experimental and GW results 
from Ref.~\citenum{shishkin_accurate_2007}.$^{(*)}$ The 
zero-point-renormalization~\cite{cardona_isotope_2005} (-0.16 eV) has 
been subtracted to the experimental gap~\cite{shishkin_accurate_2007} 
(3.44 eV) to have a meaningful comparison with the calculations.}
\label{fig:ki_zno_bands}
\end{figure*}

\noindent \textit{\bf Zinc oxide:} ZnO is a transition metal oxide which at ambient conditions
crystallizes in a hexagonal wurtzite structure. It is a well studied insulator with potential
applications in, e.g., transparent electrodes, light-emitting diodes, and solar 
cells~\cite{look_p-type_2004, look_future_2004, ozgur_comprehensive_2005,look_progress_2006}.
It is also know to be a challenging system for Green's function theory~\cite{shih_quasiparticle_2010,samsonidze_insights_2014}
and thus represents a more stringent test case for the KC functionals. 
In Fig.~\ref{fig:ki_zno_bands} the ZnO band structure calculated at different
levels of theory is shown together with experimental data. The bands around the
gap are dominated by oxygen $2p$ states in the valence and Zn $4s$ states in the
conduction with some contribution from O $2p$ and $2s$. At LDA level the band gap
is dramatically underestimated (see the table in Fig.~\ref{fig:ki_zno_bands}) 
compared to the experimental value. This underestimation is even more severe than 
in semiconductors with similar electronic structure and band gap, like e.g. GaN, 
and has been related to the O $p$ --- Zn $d$ repulsion and 
hybridization~\cite{wei_role_1988, lim_angle-resolved_2012}. In fact, at LDA level
the bands coming from Zn $d$ states lie below the O $2p$ valence bands, but are
too high in energy, resulting in upwards repulsion of the valence band maximum
states, and in an exaggerated reduction of the band gap~\cite{lim_angle-resolved_2012}.
The HSE functional pushes the $d$ states lower in energy and opens up the band gap 
(as already seen for GaAs) achieving a better agreement with experimental values. 
The KI functional moves in the same direction and further reduces the discrepancies
with experiments, providing an overall satisfactory description of the electronic 
structure. The KI performance with an error as small as 0.02 eV on the band gap
[after the zero point renormalization energy~\cite{cardona_isotope_2005}
(-0.16 eV) has been subtracted to the experimental band gap~\cite{shishkin_accurate_2007}
(3.44 eV)] and smaller than 1 eV on the $d$ band position is in line with that of 
self-consistent GW plus vertex correction in the screened Coulomb interaction.

%%%%%%%%%%%%%%%%%% ZnO Band Structure %%%%%%%%%%%%%
%%%%%%%%%%%%%%%%%%%%%%%%%%%%%%%%%%%%%%%%%%%%%%%%%%%%%%%%%%%%%%%%%%%%5

It is worth mentioning here that for the KI calculation for ZnO we used projected
Wannier functions as approximated variational orbitals. At variance with MLWFs, 
no minimization of the quadratic spread is performed in this case. The Wannier 
functions are uniquely defined by the projection onto atomic-like orbitals and, 
for the empty manifold, by the disentanglement procedure.
We found that the minimization of the quadratic spread leads to a mixing of O $2p$ 
and Zn $3d$ states with deeper ones and this deteriorate the quality of the results 
(the band gap turns out to be overestimated and the $d$ states are pushed too low in 
energy). While the KS Hamiltonian depends only on the charge density and is therefore
not affected by this unitary mixing, the orbital-density dependent part of the KI 
Hamiltonian is instead sensitive to the particular choice of the localized manifold;
the unconstrained mixing of the valence Bloch states with very different energies is
detrimental. The important question of which set of localized orbitals are the most 
suitable for the correction of the DFT Hamiltonian is not restricted to the KC functionals
but is relevant, and has indeed been discussed, also in the context of the Perdew-Zunger 
self-interaction-correction scheme~\cite{stengel_self-interaction_2008}, the generalized 
transition state method~\cite{anisimov_transition_2005, ma_using_2016}, the localized 
orbital scaling correction to approximate DFT~\cite{li_localized_2018} and the DFT+U 
method for predicting band gaps~\cite{kirchner-hall_extensive_2021}.
In principle the variational property of the KC functionals can be used to verify which
set of Wannier functions --- projected or maximally localized --- is the most energetically
favorable. This important point will be addressed in future work; here, we just highlight
the evidence that in complex systems, where an undesired mixing between state with very
different energies might be driven by the quadratic spread minimization, projected Wannier
functions seem to provide a better choice for the localized manifold.

%\lipsum[10-15]

\section{Summary and conclusions}
We have developed, tested, and described a novel and efficient implementation of KC functional
for periodic systems (but also readily applicable to finite ones) using Wannier functions as
approximated variational orbitals and a linear response approach for the efficient evaluation
of the screening parameters based on DFPT. Using the translational property of the Wannier functions
, we have shown how to recast a problem whose natural dimension is that of a supercell, into
a primitive cell problem plus a sampling of the primitive-cell Brillouin zone.
All this leads to the decomposition of the problem into smaller and independent ones
and thus to a sensible reduction of the computational cost and complexity.
We have showcased its use to compute the band structure of few prototypical systems 
ranging from small gap semiconductors to a wide-gap insulator and demonstrated that
the present implementation provides the same result as a straightforward supercell
implementation, but at a greatly reduced computational cost, thus making the KC
calculation more robust and user-friendly. The main results of Secs.~(\ref{sec:screen_coeff}) 
and (\ref{sec:ki_ham}) have been implemented as a post-processing of the PWSCF packages of the \QE{} 
distribution~\cite{giannozzi_advanced_2017, giannozzi_quantum_2009}, and of the
Wannier90 code~\cite{pizzi_wannier90_2020}, two open-source and widely used 
electronic structure tools. The KI 
implementation presented here is part of the official \QE{} distribution. It is 
hosted at the \QE{} \href{https://gitlab.com/QEF/q-e}{gitlab} repository under 
the name ``KCW'' and has been distributed with the official release $7.1$ of \QE{}.
The data used to produce the results of this work are available at the Materials
Cloud Archive.~\cite{MC}

\section{Acknowledgments}
This work was supported by the Swiss National Science Foundation (SNSF) trough its National 
Centre of Competence in Research (NCCR) MARVEL and the grant No. 200021-179138.

\bibliography{biblio}

\end{document}

\section{KI matrix elements}
For clarity and completeness, we provide here additional details on the KI matrix elements and on the derivation of Eqs.~(20) and (22) in the main text.
%Eqs.~(\ref{eq:ki-ham-K-occ}) and (\ref{eq:ki-ham-K-emp})
Let us start with the definition of the KI matrix element on two Wannier functions $| \bR i \rangle$ and $|\bzero j \rangle$: 
\begin{align}
    \Delta H^{\rm KI(2)}_{ij}(\bR) & = \me{\bR i}{\hat{\mathcal{V}}^{\rm KI(2)}_{\bzero j}}{\bzero j} \nn \\
    &= \int d\br \omega^*_{\bR i}(\br) \mathcal{V}^{\rm KI(2)}_{\bzero j}(\br) \omega_{\bzero j}(\br)
\end{align}
From a computational point of view it is convenient to first evaluate the product of the two Wannier functions in real space and then perform the integral with the KI potential in reciprocal space. Using the definition of Wannier functions in Eq.~(8) of the main text,
%Eq.~(\ref{eq:MLWF_def}),
the product of Wannier functions is given by 
\begin{align}
   \omega^*_{\bR i}(\br) \omega_{\bzero j}(\br) = & \frac{1}{N_{\bk}N_{\bk'}}\sum_{\bk \bk'} e^{i\bk \cdot \bR} \tilde{\psi}^*_{\bk i}(\br) \tilde{\psi}_{\bk' j}(\br) \nn \\
   & = \frac{1}{N_{\bk}N_{\bq}}\sum_{\bk \bq} e^{i\bk \cdot \bR} e^{i\bq \cdot \br} \tilde{u}^*_{\bk i}(\br) \tilde{u}_{\bk+\bq j}(\br) \nn \\
   & = \frac{1}{N_{\bk}}\sum_{\bk} e^{i\bk \cdot \bR} \frac{1}{N_{\bq}}\sum_{\bq} e^{i\bq \cdot \br} \rho_{\bk, \bk+\bq}^{ij}(\br),
   \label{eq_app:HR}
\end{align}
while the KI potential is given by a purely scalar term $-\frac{1}{2}\Delta^{\rm KI(2)}_{\bzero j}$ plus an $\br$-dependent contribution (only for empty states)~\cite{borghi_koopmans-compliant_2014} that can be decomposed into a sum of monochromatic terms:
\begin{align}
    \mathcal{V}^{\rm KI(2)}_{\bzero j}(\br) = -\frac{1}{2}\Delta^{\rm KI(2)}_{\bzero j} + (1-f_j) \sum_{\bq} e^{i \bq \cdot \br} V_{{\rm pert}, \bq}^{\bzero j}(\br)
    \label{eq_app:Vki}
\end{align}
Combining Eq.~\ref{eq_app:HR} and~\ref{eq_app:Vki} and using the orthogonality property of the Wannier functions $\langle \bR i | \bzero j \rangle = \delta_{\bR \bzero} \delta_{ij}$, we get
\begin{align}
    \Delta H^{\rm KI(2)}_{ij}(\bR) & = -\frac{1}{2}\Delta^{\rm KI(2)}_{\bzero j} \delta_{\bR \bzero} \delta_{ij} + \Delta H^{\rm KI(2)}_{ij, {\rm \br}}(\bR)
\end{align}
where $\Delta^{\rm KI(2)}_{\bzero j}$ is the scalar term and it is simply given by the self Hartree-exchange-correlation of the Wannier orbital density
\begin{align}
    \Delta^{\rm KI(2)}_{\bzero j} & = \langle \rho_{\bzero j} | \left[ f_{\rm Hxc} \right] | \rho_{\bzero j} \rangle = \langle \rho_{\bzero n} | V^{\bzero n}_{\rm pert} \rangle \nn \\
    & = \frac{1}{N_{\bq}}\sum_{\bq} \langle {\rho^{\bzero j}_{\bq}} | V^{\bzero j}_{{\rm pert}, \bq} \rangle.
\end{align}
and $\Delta H^{\rm KI(2)}_{ij, {\rm r}}(\bR)$ is the contribution from the $\br$-dependent part of the potential:
%\begin{widetext}
\begin{align}
    \Delta H^{\rm KI(2)}_{ij,{\rm \br}}(\bR) & = (1-f_j) \frac{1}{N_{\bk}}\sum_{\bk} e^{i\bk \cdot \bR} \frac{1}{N^2_{\bq}}\sum_{\bq \bq'} \int d\br e^{i(\bq +\bq') \cdot \br} V^{\bzero j}_{{\rm pert}, \bq}(\br) \rho_{\bk, \bk+\bq'}^{ij}(\br) \nn \\
    & = (1-f_j) \frac{1}{N_{\bk}}\sum_{\bk} e^{i\bk \cdot \bR} \frac{1}{N^2_{\bq}}\sum_{\bq \bq'}\sum_{\bg \bg'} \int d\br e^{i (\bq + \bq') \cdot \br} e^{i (\bg + \bg') \cdot \br} V^{\bzero j}_{{\rm pert}, \bq}(\bg) \rho_{\bk, \bk+\bq'}^{ij}(\bg') \nn \\
    & = (1-f_j) \frac{1}{N_{\bk}}\sum_{\bk} e^{i\bk \cdot \bR} \left\{ \frac{1}{N_{\bq}}\sum_{\bq \bg} V^{\bzero j}_{{\rm pert}, \bq}(\bg) \rho^{ij}_{\bk, \bk-\bq}(-\bg) \right\} \quad \quad [\bq \rightarrow -\bq, \bg \rightarrow -\bg] \nn \\
    & = (1-f_j) \frac{1}{N_{\bk}}\sum_{\bk} e^{i\bk \cdot \bR} \left\{ \frac{1}{N_{\bq}} \sum_{\bq \bg} V^{\bzero j}_{{\rm pert}, -\bq}(-\bg) \rho^{ij}_{\bk, \bk+\bq}(\bg) \right\} \nn \\
    & = \frac{1}{N_{\bk}}\sum_{\bk} e^{i\bk \cdot \bR} \left\{ (1-f_j) \frac{1}{N_{\bq}}\sum_{\bq \bg} [V^{\bzero j}_{{\rm pert}, \bq}(\bg)]^* \rho^{ij}_{\bk, \bk+\bq}(\bg) \right\}.
\end{align}
%\end{widetext}
The term inside the braces can be readily identified with the KI Hamiltonian at the given $\bk$-point and is the final results for the KI matrix elements reported in 
Sec.~3.1.2. From a computational point of view, the product of the (periodic part of the) Wannier functions $\rho^{ij}_{\bk, \bk+\bq}$ is performed in real space and then Fourier-transformed to the reciprocal space where the scalar product with the KI potential $V^{\bzero j}_{{\rm pert}, \bq}$ is performed. Each independent $\bq$-contribution is then summed up to give the desired matrix element.

%%%%%%%%%%%%%%%%%% GaAs Band Structure %%%%%%%%%%%%%
 \begin{figure*}[t]
    \begin{subfigure}{}
        \includegraphics[width=0.45\textwidth]{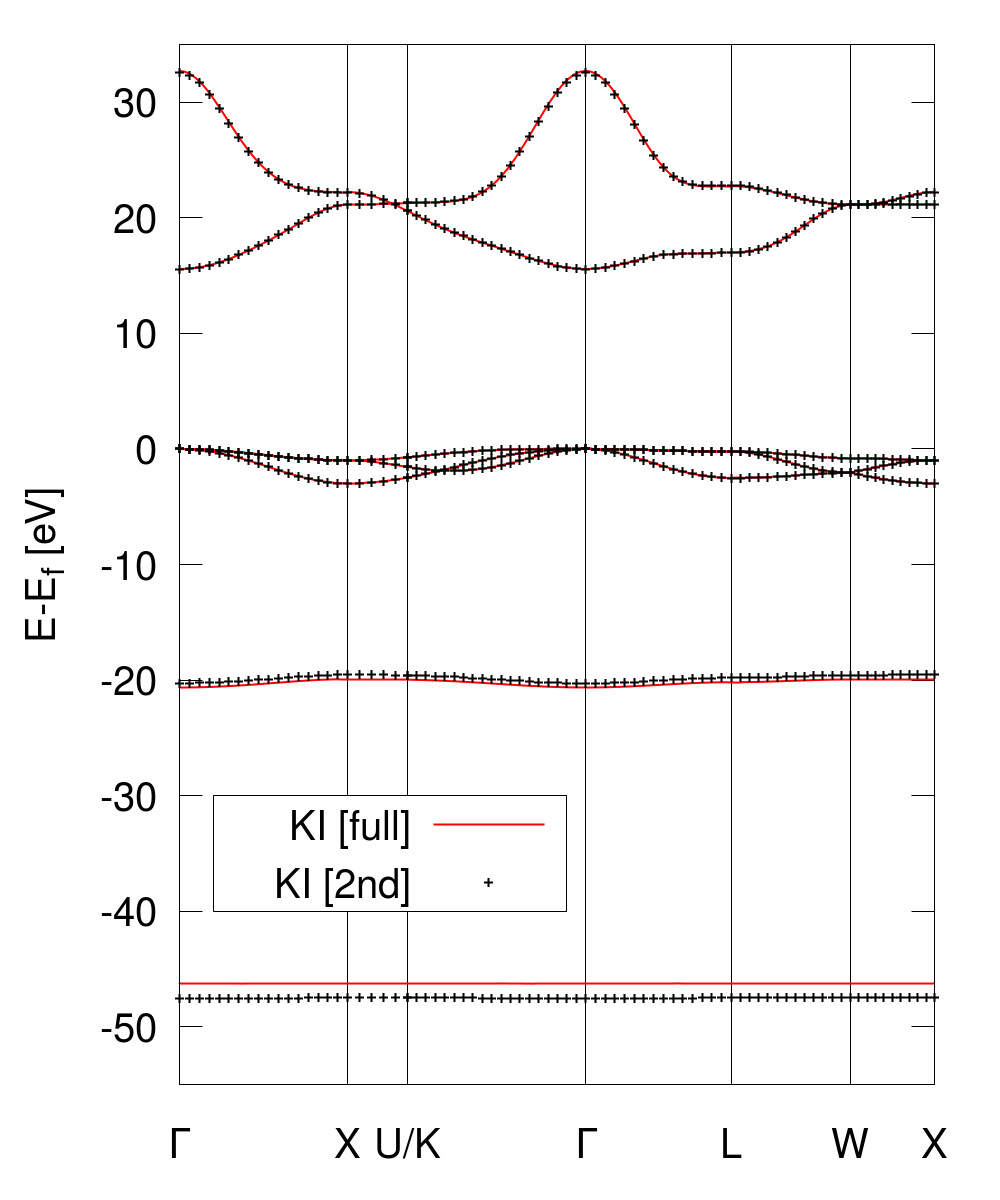}
        \includegraphics[width=0.45\textwidth]{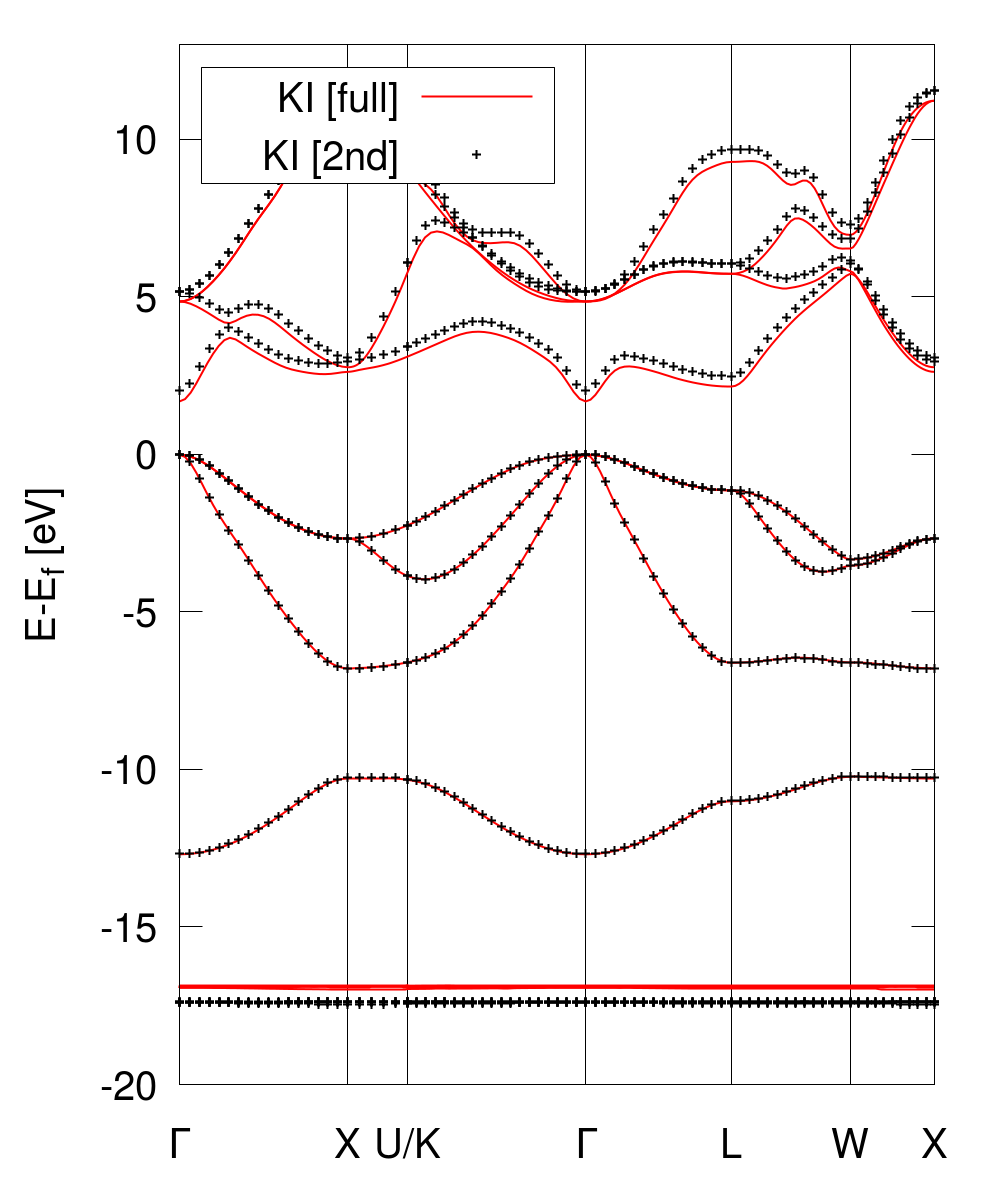}
    \end{subfigure}
%    \begin{subfigure}{} 
%        \renewcommand\tabularxcolumn[1]{m{#1}}
%        \renewcommand\arraystretch{1.3}
%        \setlength\tabcolsep{2pt}
%    \begin{tabularx}{\linewidth}{*{7}{>{\centering\arraybackslash}X}}
%        \hline
%        \hline
%        & Gap & HSE & GW$_0$ & scG$\tilde{\rm W}$ & KI & (Exp.) \\
%        \hline
%%        E$_{\rm gap}$(eV) & 8.87 & 11.61 & 13.96 &  14.5 &  15.28 & 15.35$^{(*)}$\\
%        $\langle \varepsilon \rangle_{\rm{F}_{2s}}$(eV) & -19.06 & -20.7 & -24.8$^{(\dagger)}$ &  -- &  -19.4 & -23.9 \\
%        $\langle \varepsilon \rangle_{\rm{Li}_{1s}}$(eV) & -39.6 & -42.5 & -47.2$^{(\dagger)}$ &  -- &  -46.6 & -49.8 \\
%        \hline
%    \end{tabularx}
%    \end{subfigure}
\caption{KI Band structure of LiF (left panel) and GaAs (right panel) calculated. Solid red lines are the results from a full KI calculation as described in Ref.~\citenum{de_gennaro_blochs_2021}, black cross are the results of an equivalent calculation within the present implementation and approximation.}
\label{fig:PBE}
\end{figure*}
%%%%%%%%%%%%%%%%%%%%%%%%%%%%%%%%%%%%%%%%%%%%%%%%%%%%%%%%%%%%%%%%%%%%5

\section{Full versus 2$^{\rm nd}$ order approximation}
In this section we provide additional tests for the second order approximation adopted in this work for the KI energy and potential corrections. As a reference we take a full KI calculation performed as detailed in Ref.~\citenum{nguyen_koopmans-compliant_2018} and Ref.~\citenum{de_gennaro_blochs_2021} where i) the screening coefficients are computed with a finite difference approach removing/adding one electron to the system (and not just a tiny fraction as in the 2$^{\rm nd}$ order approach presented here), and ii) the KI energy and potential corrections are not approximated to 2$^{\rm nd}$ order. We instead resort to the approximation of using MLWFs as variational orbitals also for the SC calculations presented here.

In Fig.~\ref{fig:PBE} we show a comparison between full KI band structure calculations done in a super-cell plus unfolding implementation~\cite{de_gennaro_blochs_2021} and the present approach for LiF and GaAs. We used the same computational set-up used in Ref.~\citenum{de_gennaro_blochs_2021}, i.e. a 4$\times$4$\times$4 mesh for the sampling of the Brilloiun zone to match the dimension of the supercell used in Ref.~\citenum{de_gennaro_blochs_2021}, the same lattice constants, base functional (PBE) and kinetic-energy cutoff for the expansion of the wave functions. For GaAs the smooth interpolation method presented in Ref.~\citenum{de_gennaro_blochs_2021} is also used to improve the quality of the band interpolation and to have a fair comparison with the reference results from Ref.~\citenum{de_gennaro_blochs_2021}.
For LiF we have an overall very good agreement between the two calculations; the valence and conduction bands are in perfect agreement with the reference full KI calculation, both in terms of energy position and shape. For the F $2s$ and Li $1s$ bands, lying at about -20 eV and -47 eV with respect to the top of the valence band, we observe a difference of 0.3 eV (1.5 \%) and 1.3 eV (2.8 \%), respectively. 
For GaAs the differences are more marked with a 0.3 eV difference in the band gap and 0.5 eV in the position of the Ga $d$ bands.  With respect to experiment, this leads to a slightly worse agreement for the band gap and a slightly better one for the Ga $d$ energy.  

%%%%%%%%%%%%%%%%%% LiF Band Structure %%%%%%%%%%%%%
 \begin{figure*}[t]
    \begin{subfigure}{}
        \includegraphics[width=0.45\textwidth]{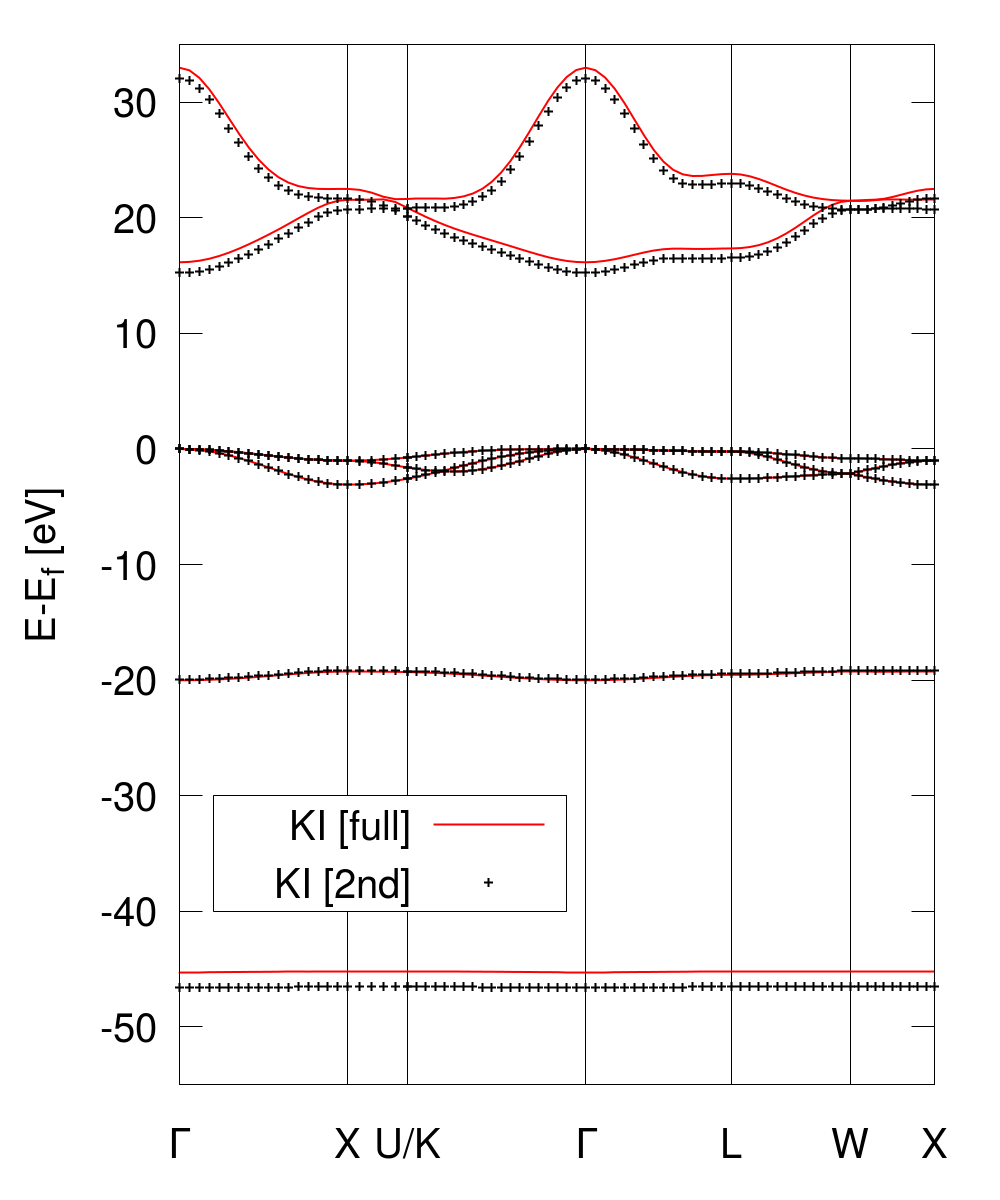}
        \includegraphics[width=0.45\textwidth]{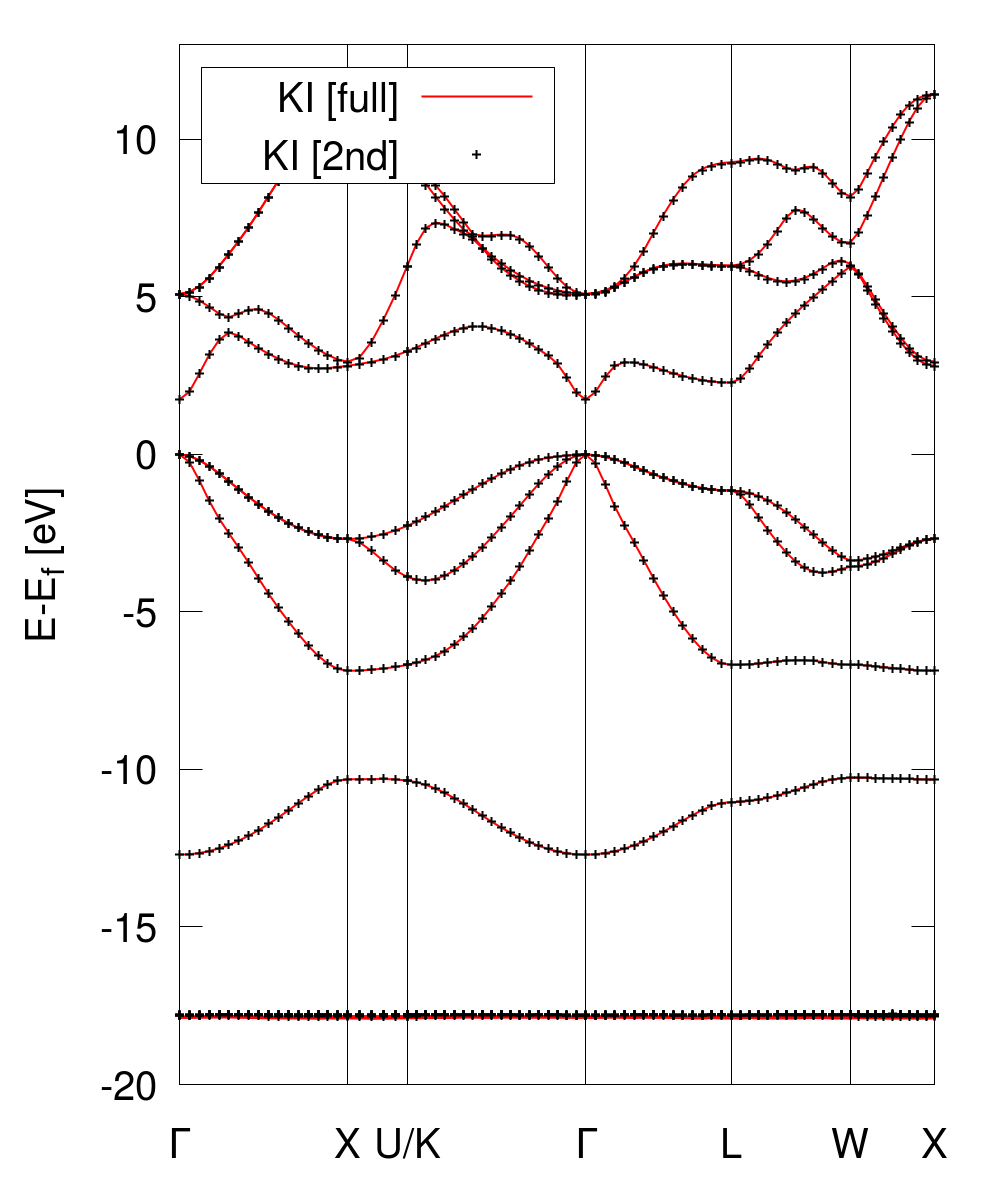}
    \end{subfigure}
%    \begin{subfigure}{} 
%        \renewcommand\tabularxcolumn[1]{m{#1}}
%        \renewcommand\arraystretch{1.3}
%        \setlength\tabcolsep{2pt}
%    \begin{tabularx}{\linewidth}{*{7}{>{\centering\arraybackslash}X}}
%        \hline
%        \hline
%        & Gap & HSE & GW$_0$ & scG$\tilde{\rm W}$ & KI & (Exp.) \\
%        \hline
%%        E$_{\rm gap}$(eV) & 8.87 & 11.61 & 13.96 &  14.5 &  15.28 & 15.35$^{(*)}$\\
%        $\langle \varepsilon \rangle_{\rm{F}_{2s}}$(eV) & -19.06 & -20.7 & -24.8$^{(\dagger)}$ &  -- &  -19.4 & -23.9 \\
%        $\langle \varepsilon \rangle_{\rm{Li}_{1s}}$(eV) & -39.6 & -42.5 & -47.2$^{(\dagger)}$ &  -- &  -46.6 & -49.8 \\
%        \hline
%    \end{tabularx}
%    \end{subfigure}
\caption{As in Fig.~\ref{fig:PBE} but using LDA as the base functional.}
\label{fig:LDA}
\end{figure*}
%%%%%%%%%%%%%%%%%%%%%%%%%%%%%%%%%%%%%%%%%%%%%%%%%%%%%%%%%%%%%%%%%%%%5

As an additional check we performed the same comparison as in Fig.~\ref{fig:PBE}, but now using the LDA as the base functional (the same used in the main text). The comparison is shown in Fig.~\ref{fig:LDA}. For GaAs the agreement between the full and the 2$^{\rm nd}$ order approximation is almost perfect, as was already observed for the density of state in Fig. 2 of the main text. At the same time we observe a slightly worse agreement for LiF, in particular the band gap is underestimated by 0.8 eV (5\%) compared to the full KI results. 
A summary of the energy positions of selected bands computed at full and 2$^{\rm nd}$ order KI for both LiF and GaAs is given in Tab.~\ref{tab:comparison}. The minor differences between the KI@LDA values reported here with respect to those in the main text (compare Tab~\ref{tab:comparison} with tables under Fig.~3 and Fig.~4 in the main text) are due to the different $\mathbf{k}$-mesh (4$\times$4$\times$4 here vs 6$\times$6$\times$6 in the main text) and, for GaAs band gap, to the spin-orbit coupling correction (-0.1 eV) applied to the results in the main text (not applied here).
This analysis points to a non trivial dependence of the quality of the 2$^{\rm nd}$ order approximation on the system and on the base functional which requires further investigation to be deeply understood and possibly corrected toward a better agreement with the full KI method. Work in this direction is under way.

\begin{table}[t]
\begin{tabular}{cccccccc}
\multicolumn{1}{l}{}                        & \multicolumn{1}{l}{}      & \multicolumn{3}{c}{LDA}                                                  & \multicolumn{3}{c}{PBE}                                            \\
\multicolumn{1}{l}{}                        & \multicolumn{1}{l}{}      &        $E_g$               & $\langle\varepsilon_{\rm{F}_{2s}}\rangle$ &  $\langle \varepsilon_{\rm{Li}_{1s}} \rangle $ &        $E_g$              & $\langle \varepsilon_{\rm{F}_{2s}} \rangle $ &  $\langle \varepsilon_{\rm{Li}_{1s}} \rangle$   \\ \hline
\multicolumn{1}{|c|}{\multirow{2}{*}{LiF}}  & \multicolumn{1}{c|}{KI Full} & \multicolumn{1}{c|}{16.16} & \multicolumn{1}{c|}{-19.6}  & \multicolumn{1}{c||}{-45.2}   & \multicolumn{1}{c|}{15.58} & \multicolumn{1}{c|}{-20.2} & \multicolumn{1}{c|}{-46.2} \\ \cline{2-8} 
\multicolumn{1}{|c|}{}                      & \multicolumn{1}{c|}{KI 2nd}  & \multicolumn{1}{c|}{15.25} & \multicolumn{1}{c|}{-19.5}  & \multicolumn{1}{c||}{-46.6}   & \multicolumn{1}{c|}{15.55} & \multicolumn{1}{c|}{-19.8} & \multicolumn{1}{c|}{-47.5} \\ \cline{2-8}
\multicolumn{1}{|c|}{}                      & \multicolumn{1}{c|}{KIPZ$^\dagger$} & \multicolumn{1}{c|}{ -- } & \multicolumn{1}{c|}{ -- }  & \multicolumn{1}{c||}{ -- }   & \multicolumn{1}{c|}{ 15.36 } & \multicolumn{1}{c|}{ -21.0 } & \multicolumn{1}{c|}{ -47.1 } \\ \hline
\multicolumn{1}{l}{}                        & \multicolumn{1}{l}{}      &        $E_g$               &        $W$                  &  $\langle \varepsilon_d \rangle $              &          $E_g$        &          $W$                & $\langle \varepsilon_d \rangle$      \\ \hline
\multicolumn{1}{|c|}{\multirow{2}{*}{GaAs}} & \multicolumn{1}{c|}{KI Full} & \multicolumn{1}{c|}{1.74} & \multicolumn{1}{c|}{12.7} & \multicolumn{1}{c||}{-17.8} & \multicolumn{1}{c|}{1.68} & \multicolumn{1}{c|}{12.7} & \multicolumn{1}{c|}{-16.9} \\ \cline{2-8} 
\multicolumn{1}{|c|}{}                      & \multicolumn{1}{c|}{KI 2nd}  & \multicolumn{1}{c|}{1.75} & \multicolumn{1}{c|}{12.7} & \multicolumn{1}{c||}{-17.8} & \multicolumn{1}{c|}{2.02} & \multicolumn{1}{c|}{12.7} & \multicolumn{1}{c|}{-17.4} \\ \hline
\end{tabular}
\caption{Relevant energy values for LiF and GaAs from full and 2$^{\rm nd}$ order KI calculations. The zero of the energy is set to the VBM. KIPZ values for LiF~\cite{de_gennaro_blochs_2021} are also reported for comparison.}
\label{tab:comparison}
\end{table}

\bibliography{biblio}